\documentclass[10pt, twocolumn]{IEEEtran}

\usepackage{subfigure}
\usepackage{latexsym}
\usepackage{amsmath}
\usepackage{amsfonts}
\usepackage{mathrsfs}
\usepackage{epsfig}
\usepackage{color}
\usepackage{wrapfig}
\usepackage{graphicx}

\usepackage[T1]{fontenc}
\usepackage{bezier}
\usepackage{times}
\usepackage{setspace}

\begin{document}

\newcommand{\no}{\noindent}
\newcommand{\be}{\begin{eqnarray}}
\newcommand{\ee}{\end{eqnarray}}
\newcommand{\beeq}{\begin{equation}}
\newcommand{\eeeq}{\end{equation}}
\newcommand{\beqs}{\begin{eqnarray*}}
\newcommand{\eeqs}{\end{eqnarray*}}
\newcommand{\bms}{\boldsymbol}

\newtheorem{theorem}{\bf Theorem}
\newtheorem{remark}{\bf Remark}
\newtheorem{result}{Result}
\newtheorem{observation}{\bf Observation}
\newtheorem{corollary}{\bf Corollary}
\newtheorem{definition}{Definition}
\newtheorem{lemma}{\bf Lemma}
\newtheorem{proposition}{\bf Proposition}

\title{ Diversity-Multiplexing Tradeoff of Asynchronous Cooperative Diversity
in Wireless Networks
}

\author{Shuangqing Wei}
\date{}
\maketitle 

\footnotetext[1]{This paper was presented in part at Allerton
Conf. Commun., Control, Computing, Oct. 2004. It will appear in the
November 2007 Issue of IEEE Transactions on Information Theory.
} 
\footnotetext[2]{This
paper was  supported in part by the Board of Regents of Louisiana
under grants LEQSF(2004-07)-RD-A-17 and NSF/LEQSF(2005)-PFUND-10.}
\footnotetext[3]{The author is with the Department of Electrical and
Computer Engineering,  Louisiana State University, Baton Rouge, LA
70803. E-mail: swei@ece.lsu.edu, Tel: (225)578-5536, Fax: (225)578-5200.}

\begin{abstract}

Synchronization of relay nodes is an important and critical issue in
exploiting cooperative diversity in wireless  networks. In this paper,
two asynchronous cooperative diversity schemes are proposed, namely,
distributed delay diversity and asynchronous space-time coded cooperative
diversity schemes. In terms of the overall diversity-multiplexing (DM)
tradeoff function, we show that the proposed independent coding based
distributed delay diversity and asynchronous space-time coded cooperative
diversity schemes achieve the same performance as the synchronous
space-time coded approach which requires an accurate symbol-level
timing synchronization to ensure signals arriving at the destination
from different relay nodes are perfectly synchronized. This demonstrates
diversity order is maintained even at the presence of asynchronism between
relay node. Moreover, when all relay nodes succeed in decoding the source
information, the asynchronous space-time coded approach is capable
of achieving better DM-tradeoff than synchronous schemes and performs
equivalently to transmitting information through a parallel fading channel
as far as the DM-tradeoff is concerned. Our results suggest the benefits
of fully exploiting the space-time degrees of freedom in multiple antenna
systems by employing asynchronous space-time codes even in a frequency
flat fading channel. In addition, it is shown asynchronous space-time
coded systems are   able to achieve  higher mutual information than
synchronous space-time coded systems for  any finite signal-to-noise-ratio
(SNR) when  properly selected baseband waveforms are employed.

\end{abstract}

\no {\bf \underline{Keywords:}} asynchronous space-time codes,
cooperative diversity, distributed delay diversity,
diversity-multiplexing tradeoff, relay channels.
 
\pagenumbering{arabic}

\section{Introduction}
\label{sec: introduction}

In wireless networks, treating  intermediate nodes between the
source and its destination as potential relays and utilizing these
relay nodes to improve the diversity gain has attracted considerable
attention lately and re-kindled interests in relay channels after
this problem was first tackled from the perspective of Shannon
capacity in the 70's \cite{meulen_71, cover:1979}.  One school of
works \cite{kramer_05,liangliang_04, gupta_03} follow the footsteps of
\cite{cover:1979},  where they employ block Markov superposition encoding,
random binning and successive decoding  as coding strategy. Another
line of work adopts the idea of cooperative diversity which was first
proposed in \cite{sendonaris_99_01, sendonaris_99_02} for CDMA networks,
and then extended to wireless networks with multiple sources and relays
\cite{laneman_02, laneman_03,falconer_04, stefanov_02, nabar_jsac_04,
hunter_02,janani_03}.  We are not attempting  to provide a comprehensive
review of all related works on relay channels here \cite{kramer_05},
but instead  divert our attentions to those work related with cooperative
diversity.

In this paper,   we mainly focus on two well received relaying strategies,
namely, decode-and-forward (DF) and amplify-and-forward (AF)  schemes.
Decision on which relaying strategy is adopted is subject to constraints
imposed upon relay nodes.  If nodes cannot transmit and receive at
the same time and thus work in a half-duplex mode \cite{ali_03},
the communication link in a relay channel with a single level of relay
nodes consists of two phases. In the first phase, the source broadcasts
its information to relays and its destination. During the second phase,
relays forward either re-encoded source transmissions (decode-and-forward)
or  a scaled version of received source signals (amplify-and-forward)
\cite{falconer_04}. At the destination, signals arriving over two phases
are jointly processed to improve the overall performance. Variations of
these schemes include allowing source nodes to continuously send packets
over two phases to increase the spectral efficiency \cite{nabar_jsac_04,
gamal_submitted}. As for coding strategies through which cooperative
diversity is achieved, \cite{stefanov_02} proposes to encode the source
information over two independent blocks from source to destination
and relays to destination, respectively. In \cite{hunter_02}, without
requiring relay nodes to provide feedback messages to the source, rate
compatible punctured convolutional codes (RCPC) and turbo codes are
proposed to encode over two independent blocks. Also, an extension is
made by putting multiple antennas at relay nodes to further improve the
diversity and multiplexing gain. If multiple relay nodes are considered
as virtual antennas, a space-time-coded cooperative diversity approach
is proposed in \cite{laneman_03} to jointly encode the  source signals
across successful relay nodes during the second phase.

As noted in \cite{falconer_review04}, synchronization of relay nodes is
an important and critical issue  in exploiting cooperative diversity in
wireless ad hoc and sensor networks.  However, in the existing works,
e.g., \cite{schein_isit_00, laneman_03}, it has been assumed that relay
nodes are perfectly synchronized  such that signals arriving at the
destination node from distinct relay nodes are aligned perfectly with
respect to their symbol epochs. Under this assumption,  distributed
space-time-coded cooperative diversity approach achieves diversity gains
in the  order of the number of available  transmitting nodes in a relay
network \cite{laneman_03}.

Perfect synchronization is, however, hard, if not impossible, to be
achieved in  infra-structureless wireless ad-hoc and sensor networks. In
\cite{anders_bounds03}, the issue of carrier asynchronism between the
source and relay node is addressed in terms of its impact on the lower and
upper bounds of the outage and ergodic capacity of a three-node wireless
relay channel. At the presence of time delays between relay nodes, an
extension of Alamouti space-time-block-codes (STBC) \cite{alamouti_98}
is proposed in \cite{xiaohua_iee03} to exploit spatial diversity
when time delay is only an integer number of symbol periods.  And in
\cite{yhao_03, hao_icc04}, macroscopic space-time codes are designed
to perform robust against uncertainties of relative delays between
different basestations. Without requiring the symbol synchronization, we
propose a repetition coding based distributed delay diversity scheme in
\cite{wei_ciss_04,wei_jsac04} which  achieves the same order of diversity
promised  by distributed space-time codes.  Unlike the extension  of other
approaches to the synchronization problem in distributed space-time coding
\cite{yhao_03},  the proposed system also admits a robust and easily
trainable receiver when synchronization is not present in the system.

In \cite{scaglione_03}, relay nodes perform adaptive decode-and-forward
or amplify-and-forward schemes allowing them to transmit or remain silent
depending on the received signal-to-noise-ratio (SNR). However, their
proposed schemes require intentionally increasing data symbol  period
to avoid inter-symbol-interference (ISI) caused by the asynchronous
transmission of the same source signal to different receivers, which
limits efficiency. In \cite{dana_it_03}, asynchronism caused by phase
error of channel fading variables is studied in terms of its impact on
relay network's energy efficiency in low SNR region.

To the best of our knowledge, there does not exist yet too much
work regarding the impact of symbol level asynchronism on the
performance of relay networks in a comprehensive manner.  The system
model in \cite{allerton04_asyn} is closest to what we assumed in
\cite{wei_allerton_04, wei_ccct_04} and this paper in terms of the consideration of
symbol level asynchronism. However, only AF scheme is considered in
\cite{allerton04_asyn} from the  perspective of the scaling law of
ergodic capacity.  In this paper, diversity-multiplexing (DM) tradeoff
function is adopted as a metric to compare the performance of our
proposed {\em asynchronous cooperative diversity} schemes with the
existing synchronous space-time-coded cooperative diversity strategy.
As first put forward  by Zheng and Tse in the context of multiple antenna
systems \cite{zheng_03_tradeoff}, the diversity-multiplexing tradeoff
function reveals a fundamental relationship between  diversity gain
which characterizes the asymptotic rate of decoding error approaching
zero as SNR increases, and multiplexing gain which characterizes
the asymptotic spectral efficiency in the large SNR regime. The
idea has recently been extended to relay channels \cite{laneman_03,
gamal_submitted}  and multiple access channels \cite{tse_zheng_mac}.

Without loss of generality (WLOG), we consider a relay channel
 where a source node communicates with its destination with
the help of two potential relays. Nodes are assumed to work
in a half duplex mode \cite{ali_03, dana_it_03, allerton04_asyn},
in which  no one can transmit and receive simultaneously. The
entire transmission period is divided into two phases. In the
first phase, source broadcasts while relays and destination
listen. In the second phase, source stops transmitting and relays
which succeed in decoding in the first phase forward source
messages to the destination, where received signals over the whole
period is jointly processed. Our major contributions can be
summarized as follows.

We first show the lower bound of the DM-tradeoff  for space-time-coded
cooperative diversity scheme developed  in \cite{laneman_03} is
actually the exact tradeoff function. In addition, it is shown the
overall DM-tradeoff under the decode-and-forward strategy is dominated
by a bottleneck case when no relay node succeeds in decoding the source
information correctly.

We then propose two asynchronous cooperative schemes under the
symbol-level asynchronism.  The first one is distributed delay diversity
scheme in which successful relay nodes forward source information
encoded with  the same codewords.  Consequently, an equivalent multipath
fading channel is constructed between relays and destination. When relay
codeword is independent of source codeword,  we prove that the overall
DM-tradeoff function remains unchanged compared with the synchronous
scheme, provided the MAC protocol ensures the relative delay $T_0$ between
two relay-destination links satisfies $T_0 \geq  2/B_w$, where $B_w$
is the bandwidth of baseband signals. When relay codeword are identical
with the source codeword, only when $B_w T_0$ is a positive integer,
can we reach the same conclusion as the independent case. Otherwise,
the overall DM-tradeoff is degraded.

The second  asynchronous cooperative diversity approach we propose is
more bandwidth efficient in that {\em asynchronous space-time codes}
are employed across successful relay nodes to  jointly encode the decoded
source information at the presence of asynchronism.  We first prove this
scheme achieves the same amount of overall diversity as the synchronous
one. Moreover, we demonstrate the presence of asynchronism provides us
an opportunity to fully exploit all degrees of freedom in space-time
domain, as evidenced by an improvement of the DM-tradeoff when all relay
nodes succeed in decoding in the first phase. Such an improvement is due
to the decoupling of the original multiple-input-single-output (MISO)
channel between relay nodes and  destination into an equivalent parallel
channel whose DM-tradeoff is better than that of a synchronous MISO
channel.  In addition, under certain conditions on baseband waveforms,
the mutual information of the asynchronous channel is even higher than
the synchronous channel  for any finite $\mbox{SNR}$.

It has been recently shown  in \cite{gamal_submitted} that the spectral
efficiency and DM-tradeoff for relay channels can be improved if a source
node keeps on transmitting signals over two phases and relay nodes don't
start forwarding until they collect sufficient information and energy to
perform the decoding. As a comparison, we propose a mixing approach where
the amplify-and-forward and asynchronous decode-and-forward schemes are
combined together. Such an approach  not only alleviates to some extent
the bottleneck caused by the absence of successful relay nodes, but also
yields a better DM-tradeoff  than schemes proposed \cite{gamal_submitted}
for some range of multiplexing gain even when the  source only broadcasts
in the first phase and stops its transmission in the second phase,
which is suggested not efficient in   \cite{gamal_submitted}. Our
results suggest the ultimate efficient relaying strategy should be
featuring both the non-orthogonal channel allocation as proposed
by \cite{gamal_submitted}, as well as the complete exploitation of
temporal-spatial degrees of freedom using asynchronous coding approach
as revealed in our analysis.

This paper is organized as follows. The system model of a relay channel
is introduced in Section \ref{section: system model}.  We revisit
the DM-tradeoff of the synchronous space-time-coded scheme proposed
by \cite{laneman_03} in Section \ref{section: syn. distc} and prove
their lower bound  is actually the exact value. An independent coding
based and repetition coding based distributed delay diversity schemes
and an asynchronous space-time coded cooperative diversity scheme are
proposed  in Section~\ref{section: Delay Diversity} and ~\ref{section:
asyn. coding}, respectively. Their DM-tradeoff are analyzed  and compared
against the synchronous coded approach. A mixing relaying strategy
combining DF and AF is proposed in Section~\ref{sec: bottle neck}
to resolve to certain extent the bottleneck issue which restricts the
overall DM-tradeoff for orthogonal relay channels.  Finally, we conclude
the paper in Section~\ref{section: conclusion}.

\section{System Model}
\label{section: system model}

\begin{figure}[h]
         \centerline{
            {
\includegraphics[scale=0.55]{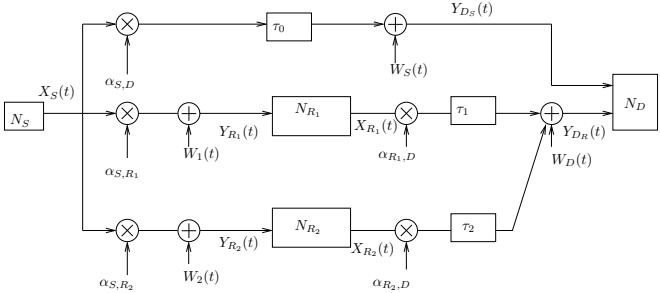}
             }
           }
         \caption{System model of an ad hoc wireless network}
         \label{system_model}
\end{figure}

To simplify analysis and reveal fundamental insights, we consider a
relay network where a source node transmits messages to its destination
node with the help of $K=2$ relays.  It is assumed relay nodes work in a
half-duplex mode, which prohibits  them  from transmitting  and receiving
at the same time \cite{ali_03}.  As assumed in \cite{laneman_03}, the
system works in two phases.  In the first phase, the source broadcasts
its transmission to its destination and potential relays. In the
second phase, the source remains silent and only those relays which
succeed in decoding the source information forward the packets  after
reprocessing.  A mathematical model of  such a network is shown in
Figure~\ref{system_model}.

After some processing of the received signal $Y_{R_k}(t), k=1,2$ from the
source node $N_S$ at the $k$th relay node $N_{R_k}$, $N_{R_k}$ transmits
the processed packets via $X_{R_k}(t)$ to the destination node $N_{D}$,
where signals from all involved paths are processed jointly.  Quasi-static
narrow-band transmission is assumed where the channel between any pair of
nodes is frequency non-selective, and the associated fading coefficients
remain unchanged  during the transmission of a whole packet, but are
independent from node to node and packet to packet.  Time delays $\{
\tau_k\}$ are introduced on each path, which incorporate the processing
time at relay nodes and  propagation delays of the whole route. More
specifically, $\tau_0$ is the delay from $N_S$ to $N_D$, and $\tau_k$
is the cumulative delay for the transmission from $N_S$ to $N_{R_k}$,
processing at $N_{R_k}$ and for transmission from $N_{R_k}$ to $N_{D}$,
for $k=1,2$. 

The noise processes $W_S(t)$, $W_D(t)$ and $W_k(t), k=1,
2$ are independent complex white Gaussian noise with two-sided power
spectral density ${\cal N}_0$. Assume signals $X_i(t), i \in \left\{S,
R_1, R_2\right\}$ share a common radio channel with complex baseband
equivalent bandwidth $[-B_w/2, B_w/2]$ and each node transmits signals
of duration $T_d$, which leads to the transmission of $L= \lfloor B_w
T_d \rfloor$ independent complex symbols over one packet. 
Define $\mbox{SNR} \stackrel{\Delta}{=}
\frac{P_s}{{\mathcal N_0} B_w} = \frac{\hat{P}_s}{{\mathcal N}_0}$,
where $P_s$ and $\hat{P}_s = P_s/B_w$ are the common continuous and
discrete time transmission power of each transmitting node, respectively
\cite{laneman_03}, which are assumed fixed.

The complex channel gain $\alpha_{i,j}$ captures the effects of
both pathloss and quasi-static fading on links between node $N_i$
and  node $N_j$, where $i \in \left\{S, R_1,  R_2 \right\}$, and $j
\in \left\{R_1, R_2, D \right\}$. Statistically, $\alpha_{i,j}$ are
modeled  as  zero mean, mutually independent complex  Gaussian
random variables with variances $\sigma^2_{i,j}$. The fading variances
are specified  using wireless path-loss models based on the network
geometry \cite{rappaport_96}. Here, it is assumed that  $ \sigma^2_{i,j}
\propto 1/d^{\mu}_{i,j}$, where $d_{i,j}$ is the distance from node $N_i$
to $N_j$, and $\mu$ is a constant whose value, as estimated from field
experiments, lies in the range $2 \leq \mu \leq 5$. Throughout this
paper, we assume $\alpha_{i,j}$ is perfectly known  at  receiver $N_j$,
but  not available to the transmitter $N_i$. Consequently, transmission
schemes exploiting transmitter side channel state information (CSI),
such as successive encoding \cite{caire_BC_2003} using dirty paper coding
approach  \cite{Costa_DPC} and power control schemes \cite{wei_ISIT2006}
, are not considered in this paper.

The two-phase transmission and half-duplex  mode of relay nodes results
in orthogonality in time between the  packet arriving at $N_D$ via the
direct path from $N_s$ and  the collection of packets arriving at $N_D$
through  different relay nodes. Note that the  orthogonality between
signals $X_{R_1}(t)$ and $X_{R_2}(t)$ is
 {\em not} assumed, which  forms the crux of the
problem. Time difference $\tau_k -\tau_0$ incorporates the processing time
of a whole packet at $N_{R_k}$ in addition to the relative propagation
delay between the $k$th relay path and the direct link. Without loss of
generality (WLOG),  $\tau_0$ is set to zero.
 Under the preceding model,  the received signals in Fig.
\ref{system_model} are specified by :  
\begin{gather}
Y_{R_k}(t) = \alpha_{S,
R_k} X_S(t) + W_k(t), \, k = 1, 2, \nonumber \\
Y_{D_s}(t) =\alpha_{S,D}
X_S\left(t\right) + W_S(t), \nonumber \\
Y_{D_R}(t) = \sum_{j \in {\mathcal D}(s)} \alpha_{j, D}
X_{j} \left(t - \tau_j \right) + W_D(t), 
\end{gather}

\no where $Y_{D_s}(t)$ and $Y_{D_R}(t)$ have no common support in time
domain, and  ${\mathcal D}(s)$ denotes  the set of relay nodes which have
successfully decoded the information from $N_s$, whose cardinality
$|{\mathcal D}(s)|$  satisfies $|{\mathcal D}(s)| \in \left\{ 0, 1,
2 \right\}$.

\section{Diversity-Multiplexing Tradeoff}
\label{section: information theory}

\subsection{Synchronous Distributed Space-time-Coded Cooperative Diversity}
\label{section: syn. distc}

The DM-tradeoff of the distributed space-time-coded cooperative relaying
proposed in \cite{laneman_03} is revisited in this section. To study
DM-tradeoff function, the source transmission rate $R$(bits/second/Hz)
needs to be parameterized as a function of the  transmission $\mbox{SNR}$
as follows \cite{laneman_03}, \beeq \label{eq: rate R} R \left(\mbox{SNR}
\right) = r \log \left(1 + \mbox{SNR} \sigma^2_{S,D} \right), \eeeq

\no where
$0<r \leq  1$ characterizes the spectral efficiency normalized by
the direct link channel capacity, which illustrates how fast the
source data rate varies with respect to $\mbox{SNR}$ and is
defined as the multiplexing gain in \cite{zheng_03_tradeoff}, i.e.
\[ r = \lim_{\small \mbox{SNR} \rightarrow \infty} \frac{R \left(\mbox{SNR}
\right)}{\log
\mbox{SNR}}.
\]

\no  A fundamental figure introduced in \cite{zheng_03_tradeoff} is the
diversity-multiplexing tradeoff which illuminates the relationship between
the reliability of data transmissions in terms of diversity gain, and
the spectral efficiency in terms of multiplexing gain. This relationship
can be characterized by mapping the diversity gain as a function of $r$,
i.e. $d(r)$, where $d(r)$ is the diversity gain and  defined by \beeq
\label{eq: def. of diversity gain} d(r) = \lim_{ \mbox{SNR} \rightarrow
\infty} - \frac{ \log \left( \mbox{Pr} \left[ I < R \left( \mbox{SNR}
\right) \right] \right)}{\log \mbox{SNR}}, \eeeq

\no where $I$ is the mutual information between the source and its destination node.

Laneman and Wornell  developed lower and upper bounds of  this tradeoff
function for space-time-coded cooperative diversity scheme by assuming
perfect symbol-level synchronization \cite{laneman_03}. Denote
$d_{stc}(r)$ as the corresponding tradeoff function. The bounds of 
$d_{stc}(r)$ are   \beeq \label{eq: laneman IT_03 result}
(K+1)(1-2r) \leq d_{stc} (r) \leq (K+1) \left( 1 - \frac{K}{K+1} \cdot
2 r \right), \eeeq

\no where  $K+1$ denotes the total number of potential transmitting nodes
in the network. In this paper, we have $K+1=3$ for a four-node network.
When $d_{stc}(r)$ is computed using the definition of (\ref{eq: def. of
diversity gain}), $\mbox{Pr} \left[ I < R \left( \mbox{SNR} \right)
\right]$ is the outage probability that the mutual information of an
equivalent channel between the source and its destination is below the
parameterized spectral efficiency  $R$ when all possible outcomes of
relays decoding source signals are counted. Next, we  show the lower
bound  $3-6r$  in (\ref{eq: laneman IT_03 result}) is actually tight.

\begin{theorem} \label{theorem1} The lower bound of the
diversity-multiplexing tradeoff for the synchronous space-time-coded
cooperative diversity developed in \cite{laneman_03} is tight,
i.e. $d_{stc}(r) = (K+1) (1 -2 r)$.  \end{theorem}

\begin{proof}
For comparison purpose,  similar definitions as in
\cite{laneman_03} are adopted in the sequel. It will be shown
below that a bottleneck case dominates the overall diversity
order $d_{stc}(r)$ and thus leads to the desired result.

Suppose
identically and independently distributed (i.i.d) circularly
symmetric, complex Gaussian codebooks are employed by the source
and all successful relay nodes. Conditioned on the decoding set
$\mathcal{D}(s)$, the mutual information $I_{stc}$ between $N_S$
and $N_D$ of the distributed space-time-coded scheme with perfect
synchronization is \cite[Eq. (18)]{laneman_03}
 \begin{gather} \label{eq: STC mutual infor}
I_{stc} = \frac{1}{2} \log  \left( 1 + \frac{2}{K+1}
\mbox{SNR} \left|\alpha_{S,D} \right|^2 \right) + \nonumber \\ 
\frac{1}{2} \log
\left( 1 + \frac{2}{K+1} \mbox{SNR} \sum_{R_k \in {\mathcal D(s)}}
 \left| \alpha_{R_k, D} \right|^2   \right),
\end{gather}

\no where $2/(K+1)$ is a normalization factor introduced to make a
fair comparison with the non-cooperative scheme and the factor
$1/2$ in front of $\log$-functions  is due to the encoding over two independent
blocks.

The outage probability can be calculated based on the total probability
law \beeq \label{eq: def. of the outage prob} \mbox{Pr} \left[ I_{stc}
< R \right] = \sum_{{\mathcal D}(s) } \mbox{Pr} \left[ {\mathcal D}
(s) \right] \mbox{Pr} \left[ I_{stc} < R | {\mathcal D}(s) \right], \eeeq

\no where the probability of the decoding set  is 
\beeq \label{eq: decoding set prob}
\mbox{Pr} \left[ {\mathcal D} (s) \right] = \prod_{R_k \in
{\mathcal D}(s) } \mbox{Pr} \left[ I_{S,R_k} \geq  R \right]
\times \prod_{R_j \not\in {\mathcal D}(s) } \mbox{Pr} \left[
I_{S,R_j} <  R  \right], \eeeq

\no and $I_{S,R_j}$ is the mutual information between $N_s$ and
$N_{R_j}$ using i.i.d complex Gaussian codebooks, and is given by
\beeq \label{eq: mutual of S and R} I_{S,R_j} = \frac{1}{2} \log
\left( 1 + \frac{2}{K+1} \mbox{ SNR} \left| \alpha_{S,R_j} \right|^2
\right). \eeeq

\no In order to derive the overall tradeoff function $d_{stc}(r)$, we
need to study the asymptotic behavior of all sum terms in (\ref{eq:
def. of the outage prob}) where $R$ should be replaced by (\ref{eq:
rate R}). However in \cite{laneman_03}, the bounds of $d_{stc}(r)$ in
(\ref{eq: laneman IT_03 result}) are  developed by first fixing $R$
in order to obtain an asymptotic equivalence form of $\mbox{Pr} \left[
I_{stc} < R \right]$ and then substituting  the rate $R$ with  $R({\mbox
SNR})$. This approach conceals the dominance of the worst situation when
all relay nodes fail in decoding  source messages, which consequently
drags down the overall diversity order in an overwhelming manner.
This point will be made more clearly through our asymptotic analysis.

Consider first the outage probability  $\mbox{Pr} \left[ I_{S,R_j}
<  R  \right] $ for large $\mbox{SNR}$: 
\begin{gather} \label{eq: prob. of out of decoding set} 
\mbox{Pr} \left[ I_{S,R_j} <  R \right] = 
\mbox{Pr} \left[ \frac{1}{2} \log \left( 1+ \frac{2}{K+1} \mbox{SNR}
|\alpha_{S,R_j}|^2 \right) < \right. \nonumber \\
\left. r \log \left(1 + \mbox{SNR}
\sigma^2_{S,D} \right)
\right] \nonumber \\
 \sim  \mbox{Pr} \left[ |\alpha_{S,R_j}|^2 < \mbox{SNR}^{2r-1}
\frac{\left( \sigma_{S,D}^2
\right)^{2r}}{2/(K+1)} \right] \nonumber \\
 =  1 - \exp\left\{ -\lambda_{S,R_j} \mbox{SNR}^{2r-1} c_0
\right\}, \end{gather}

\no where ``$\sim$" is the symbol representing  
an asymptotic equivalence at large $\mbox{SNR}$
\cite{karl_aysmptotic}, i.e. as $\mbox{SNR} \rightarrow \infty$,
$f(\mbox{SNR}) \sim g(\mbox{SNR})$ $\Rightarrow \lim_{{
\mbox{SNR}} \rightarrow \infty} f(\mbox{SNR})/g(\mbox{SNR}) =1 $.
With  $c_0 =
\frac{\left( \sigma_{S,D}^2 \right)^{2r}}{2/(K+1)}$,
 the second equality is because $|\alpha_{i,j}|^2$ is exponentially distributed
with parameter $\lambda_{i,j} = 1/\sigma_{i,j}^2$.  It can be seen
from (\ref{eq: prob. of out of decoding set}) that if $r \geq
1/2$, the  probability of no  successful relay nodes, i.e.
$|{\mathcal D}(s)| =0$, is in an  order of a non-zero constant for
large $\mbox{SNR}$. In addition, the conditional overall outage
probability given $|{\mathcal D}(s)| =0$ can be determined
similarly by
 \beeq \label{eq: conditional whole link witihout
relay} \mbox{Pr}\left[ I_{stc} < R | |{\mathcal D}(s)| = 0 \right]
\sim 1 - \exp\left\{ -\lambda_{S,D} \mbox{SNR}^{2r-1} c_0 \right\}
, \eeeq

\no which is also in the order of a non-zero constant when $r \geq
1/2$. Therefore, if $r\geq 1/2$, the overall outage probability
$\mbox{Pr} \left[ I_{stc} < R \right]$ is dominated by a non-zero
and non-vanishing term as   $\mbox{SNR} \rightarrow \infty$ which
is of no interest to our investigation of the DM-tradeoff.
Actually, such limitation imposed on multiplexing gain is due to
our restriction of letting source and relay nodes work in the half
duplex mode. Recently, cooperative diversity schemes addressing
this  half duplex limitation are proposed  in
\cite{gamal_submitted}. In Section~\ref{sec: bottle neck}, we will
make comparisons between our proposed strategies and those in
\cite{gamal_submitted} to illustrate  benefits of exploiting asynchronism.
For schemes proposed subsequently in this
paper, we only consider  multiplexing gains  $r \in [0, 1/2)$.
Under such condition and  $e^{x} \sim 1+x$ for $x \rightarrow 0$,
we obtain
 \beeq
\mbox{Pr} \left[ I_{S,R_j} <  R  \right] \sim \lambda_{S,R_j} c_0
\mbox{SNR}^{-(1-2r)},  \eeeq

\no for $0 \leq r <1/2, \, j=1,2$.

 Thus, the probability of the decoding set $\mathcal{D}(s)$ is
\beeq \label{eq: asymptotic of decoding set} \mbox{Pr} \left[
{\mathcal D}(s) \right]  \sim \left[ c_0 \mbox{SNR}^{-(1-2r)}
\right]^{K-|{\mathcal D}(s)|}
 \prod_{j\not\in {\mathcal D}(s)} \lambda_{S,R_j}, \eeeq

\no where $|{\mathcal D}(s)| \in \left\{ 0, 1, 2
 \right\}$.

Combining (\ref{eq: conditional whole link witihout relay}) and
(\ref{eq: asymptotic of decoding set}), we obtain
 \beeq \label{eq: whole link without relay} \mbox{Pr}\left[ I_{stc} < R, \,
|{\mathcal D}(s)| = 0 \right] \sim \lambda_{S,D} \prod_{j=1}^2
\lambda_{S,R_j}  c_0^3 \cdot \mbox{SNR}^{-3(1-2r)}. \eeeq

Next, we show when $|\mathcal{D}(s)|>0$, the overall diversity is
dominated by the term $3(1-2r)$, i.e.  $\mbox{SNR}^{-3(1-2r)}$
becomes the slowest vanishing  term  as $\mbox{SNR} \rightarrow
\infty$.

To simplify denotations, we define $\widetilde{
\mbox{SNR}} = \sigma_{S,D}^2 {\mbox{SNR}}$ and $
|{\tilde\alpha}_{i,j}|^2 = \frac{2/(K+1)}{\sigma_{S,D}^2} \cdot
|\alpha_{i,j}|^2$.  Random variables $|{\tilde\alpha}_{i,j}|^2$
are exponentially distributed  with parameters
$\tilde{\lambda}_{i,j}= \frac{\sigma_{S,D}^2}{2/(K+1)} \cdot
\lambda_{i,j}$. In order to study the asymptotic behavior  of the
conditional outage probability $\mbox{Pr} \left[ I_{stc} < R\, |
\, {\mathcal D}(s) \right]$ for $|{\mathcal D}(s)| > 0$, we
further normalize
 $|{\tilde\alpha}_{i,j}|^2$ by
 $\beta_{i,j}= -\frac{\log |{\tilde\alpha}_{i,j}|^2}{\log
\widetilde{\mbox{SNR}}}$
 \cite{zheng_03_tradeoff},  which yields $ \left( 1 +
\widetilde{\mbox{SNR}} |{\tilde\alpha}_{i,j}|^2 \right) \sim
\widetilde{\mbox{SNR}}^{(1-\beta_{i,j})^+} $ for large $\mbox{SNR}$, where
$(z)^+$ denotes $\max\left\{z,0 \right\}$. Thus, the conditional
outage probability given $N_{R_1} \in {\mathcal D}(s)$ is 
\begin{gather} \label{eq: outage prob. asymp. with one relay node} 
\mbox{Pr}\left[ I_{stc} < R | |{\mathcal D}(s)|=1, N_{R_1} \in {\mathcal
D}(s)  \right] \nonumber \\ 
= \mbox{Pr}  \left[ \sum_{i\in \left\{S,R_1
\right\}} \log\left( 1 + \widetilde{\mbox{SNR}}
\left|{\tilde\alpha}_{i,D} \right|^2 \right)
< 2r \log \left( 1 + \widetilde{\mbox{SNR}}  \right) \right] \nonumber \\
 \sim 
\mbox{Pr} \left[ \widetilde{\mbox{SNR}}^{\sum_{i \in \left\{S,R_1 \right\}} (1- \beta_{i,D})^{+}} < \widetilde{\mbox{SNR}}^{2r} \right] \nonumber \\
 =  \mbox{Pr} \left[ \sum_{i \in \left\{S,R_1 \right\}} (1- \beta_{i,D})^{+} < 2r   \right] \nonumber \\
 =  \int_{\underline{\beta} \in {\mathcal A}}  \left( \log
\widetilde{\mbox{SNR}} \right)^2 \prod_{k \in \{S, R_1 \}}
\widetilde{\mbox{SNR}}^{- \beta_{k,D}}
\tilde{\lambda}_{k,D}  \nonumber \\
  \exp\left\{ -\tilde{\lambda}_{k,D} \widetilde{\mbox{SNR}}^{-
\beta_{k,D}} \right\} \, d \beta_{S,D} d \beta_{R_1,D}, 
\end{gather}

\no where $\underline{\beta} = \{\beta_{i,D}\}$ and ${\mathcal A} =
\left\{\underline{\beta}: {\sum_{i \in
\left\{S,R_1 \right\}} (1- \beta_{i,D})^{+}} < 2r \right\}$, and
the last equality is yielded by integrating the joint probability
density function of the vector of $\{ \beta_{i,D} \}$ over
${\mathcal A}$.
  As shown in \cite[pp. 1079]{zheng_03_tradeoff},
we only need to consider the set
\[
\tilde{\mathcal A} = \left\{\underline{\beta}:  {\sum_{i \in
\left\{S,R_1 \right\}} (1- \beta_{i,D})^{+}} < 2r , \beta_{i,D}
\geq 0 \right\}\]

\no for the asymptotic behavior of the right hand side (RHS) of
(\ref{eq: outage prob. asymp. with one relay node}) since the term
$\exp\left\{ -\tilde{\lambda}_k \widetilde{\mbox{SNR}}^{-
\beta_{k,D}} \right\}$ decays exponentially fast for any
$\beta_{i,D} <0$ whose exclusion does not affect the diversity
order.
 Therefore, \begin{gather}
\label{eq: outage prob. without exp with one relay node} \mbox{Pr}
\left[ I_{stc} < R | {\mathcal D}(s) = \left\{ N_{R_1} \right\}
\right] \sim \int_{ \underline{\beta} \in \tilde{\mathcal A}}  \left(
\log \widetilde{\mbox{SNR}} \right)^2 \nonumber \\ \prod_{k \in \{S, R_1 \}}
\widetilde{\mbox{SNR}}^{- \beta_{k,D}} \tilde{\lambda}_{k,D} \, d
\beta_{S,D} d \beta_{R_1,D}. \end{gather}

\no As we need to obtain the asymptotic relation of all sum terms in
(\ref{eq: def. of the outage prob}),  studying an asymptotic equivalence of
$\log
\left(\mbox{Pr}\left[ I_{stc} < R | {\mathcal D}(s) = \left\{
N_{R_1} \right\} \right] \right)$ as $\log\mbox{SNR} \rightarrow
\infty$ is not sufficient to give us the desired asymptotic equivalence  for
$\mbox{Pr}\left[ I_{stc} < R | {\mathcal D}(s) = \left\{
N_{R_1} \right\} \right] $ because in general we have:
\cite[p. 38]{karl_aysmptotic} \beeq \label{eq: not equivalent
rule} \log\left( f(x) \right) \sim \log\left( g(x) \right), \, x
\rightarrow \infty \not\Rightarrow f(x) \sim g(x), \, x
\rightarrow \infty. \eeeq

\no Consequently, we need to delve into more precise asymptotic characterization
of (\ref {eq: outage prob. without exp with
one relay node}) by dividing $\tilde{\mathcal A}$ into four
non-overlapping subsets: $\tilde{\mathcal A} = \bigcup_{i=1}^{4}
\tilde{\mathcal A}_i$, where $\tilde{\mathcal A}_1 = \left\{
\beta_{S,D} \geq 1, \beta_{R_1,D} \geq 1 \right\}$,
$\tilde{\mathcal A}_2 = \left\{ \beta_{S,D} \geq 1, \right.$
$\left. 1- 2r < \beta_{R_1,D} < 1 \right\}$, $\tilde{\mathcal A}_3
= \left\{ 1-2r < \beta_{S,D}  < 1, \beta_{R_1,D} \geq  1 \right\}$
and $\tilde{\mathcal A}_4 = \left\{ 0 \leq \beta_{k, D} < 1,
\right.$ $\left. \sum_{k \in \{S, R_1 \}}\right.$ $\left. \beta_k
> 2 -2r \right\}$. As a result, the RHS of (\ref{eq: outage prob.
without exp with one relay node}) is divided into four terms each
of which is an integral over ${\mathcal A}_i, i =1, \ldots, 4$,
respectively. The asymptomatic equivalence of each term is then
studied individually leading  to Lemma~\ref{lemma1}.

\begin{lemma} \label{lemma1}
The asymptotic equivalence of the RHS of (\ref{eq: outage prob.
without exp with one relay node}) is
\begin{gather} \label{eq: asymp. conditional final of single realy node}
\int_{\underline{\beta} \in \tilde{\mathcal A}}  \left( \log
\widetilde{\mbox{SNR}} \right)^2 \prod_{k \in \{S, R_1 \}}
\widetilde{\mbox{SNR}}^{- \beta_{k,D}} \tilde{\lambda}_{k,D} \, d
\beta_{k,D}
\sim  \nonumber \\
 \left( 2 r \log \widetilde{\mbox{SNR}}\right) \left(
\widetilde{\mbox{SNR}} \right)^{-(2-2r)} \prod_{k \in \{S, R_1 \}}
\tilde{\lambda}_{k,D}. \end{gather}

\end{lemma}

\begin{proof}  See  Appendix~\ref{pf: lemma1}

\end{proof}

\no If $f \sim \phi$ and $g \sim \psi$ as $x \rightarrow x_0$, we
have $fg \sim \phi \psi$ \cite{karl_aysmptotic}. Thus, combining
(\ref{eq: asymptotic of decoding set}), (\ref{eq: outage prob.
without exp with one relay node}) and
(\ref{eq: asymp. conditional final of single realy node})
yields:

\begin{lemma} \label{lemma2}
The asymptotic equivalence   of the outage probability for
${\mathcal D}(s) = \left\{ N_{R_1} \right\}$ is
\begin{gather}
\label{eq: whole link outage with nr-1 relay node} \mbox{Pr}
\left[ I_{stc} < R,  {\mathcal D}(s) = \left\{ N_{R_1} \right\}
\right] \nonumber \\ 
\sim c_0 \tilde{\lambda}_{S,D} \tilde{\lambda}_{R_1,D}
\tilde{\lambda}_{S,R_2} \left( \widetilde{\mbox{SNR}}
\right)^{-(3-4r)} \left( 2 r \log \widetilde{\mbox{SNR}}\right) .
\end{gather}
\end{lemma}

\no It can be shown using the similar approach that 
\begin{gather}
\label{eq: whole link outage with nr-2 relay node} \mbox{Pr}
\left[ I_{stc} < R,  {\mathcal D}(s) = \left\{ N_{R_2} \right\}
\right] \nonumber \\ 
\sim c_0 \tilde{\lambda}_{S,D} \tilde{\lambda}_{R_2,D}
\tilde{\lambda}_{S,R_1} \left( \widetilde{\mbox{SNR}}
\right)^{-(3-4r)} \left( 2 r \log \widetilde{\mbox{SNR}}\right),
\end{gather}

\no which makes the following asymptotic equivalence hold, 
\begin{gather} \label{eq: general link outage with one realy node} \mbox{Pr}
\left[ I_{stc} < R, |{\mathcal D}(s)| =1  \right]  \sim \left(
\tilde{\lambda}_{S,R_1}\tilde{\lambda}_{R_2,D} +
\tilde{\lambda}_{S,R_2}\tilde{\lambda}_{R_1,D} \right)
\tilde{\lambda}_{S,D} \nonumber \\
\left( \widetilde{\mbox{SNR}} \right)^{-(3-4r)} \left( 2 r \log
\widetilde{\mbox{SNR}}\right). 
\end{gather}

The only term left in (\ref{eq: def. of the outage prob})
represents the case when two relay nodes both succeed in decoding
the source messages and then jointly encode using i.i.d complex
Gaussian codebooks independent of the source codewords.  For this
case, we obtain
\begin{lemma} \label{lemma3}
When both relay nodes are in the decoding set, the overall outage
probability has an asymptotic behavior characterized by \beeq
\label{eq: full outage prob. with 2 relay nodes} \mbox{Pr} \left[
I_{stc} < R, \; |{\mathcal D}(s)| =2   \right] \sim
 2 \prod_{k \in \left\{ S,R_1,R_2 \right\}} \tilde{\lambda}_{k,D}
\left( \widetilde{\mbox{SNR}} \right)^{-3 + 4 r}. \eeeq
\end{lemma}

\begin{proof} See  Appendix~\ref{lemma3proof}.
\end{proof}

Given the asymptotic equivalence of outage probabilities $
\mbox{Pr} \left[ I_{stc} < R, \; |{\mathcal D}(s)| =j \right]$ for
$ j  \in \{0, 1, 2 \}$  in (\ref{eq: whole link without relay}),
(\ref{eq: general link outage with one realy node}) and (\ref{eq:
full outage prob. with 2 relay nodes}), we can conclude the
overall decaying rate of $ \mbox{Pr} \left[ I_{stc} < R \right]$ towards zero is
subject to the worse case when there is no relay node
in the decoding set because $ \mbox{SNR}^{-3+6r}$ in (\ref{eq:
whole link without relay}) dominates $\mbox{SNR}^{-3+4r}$ in
(\ref{eq: general link outage with one realy node}) and (\ref{eq:
full outage prob. with 2 relay nodes}) for large $\mbox{SNR}$.
Therefore, the overall outage probability has the following
asymptotic behavior,
 \beeq  \label{eq: diversity/mutlipexling
tradeoff syn. st. coding} \mbox{Pr} \left[ I_{stc} < R \right]
\sim \prod_{k \in \left\{D, R_1, R_2 \right\} }
\tilde{\lambda}_{S,k} \left( \widetilde{\mbox{SNR}} \right)^{-3 +
6 r}, \, 0 \leq r < \frac{1}{2} ,\eeeq

\no which implies $d_{stc}(r) = 3 \left( 1- 2 r \right), \, 0 \leq
r < 1/2$.  This is the lower-bound of (\ref{eq: laneman IT_03
result}) developed in   \cite{laneman_03} for $K+1=3$. It means the worst
scenario in a cooperative diversity scheme using the
decode-and-forward strategy is when all relay nodes fail to decode
the source packets correctly and  the DM-tradeoff function under
this case becomes  the dominant one in determining the overall
DM-trade-off function $d_{stc}(r)$. This conclusion can be
extended in a straightforward manner to the case of more than 2 relay nodes
yielding \be\label{eq: extended diverstiy trade-off for syn. space-time}
d_{stc}(r) & = &  \lim_{  \mbox{SNR} \rightarrow \infty} -
\frac{ \log \left( \mbox{Pr} \left[ I < R \left( \mbox{SNR}
\right) \right] \right)}{\log \mbox{SNR}} \nonumber  \\
&  = & (K+1) ( 1 - 2 r), \, 0 \leq r < 1/2,
\ee 
\no which thus proves Theorem~\ref{theorem1}.

\end{proof}

Next, without assuming perfect synchronization between relay
nodes,  we  investigate the impact of asynchronism on the overall
diversity-multiplexing tradeoff for cooperative diversity schemes.
This asynchronism is presented in terms of non-zero relative
delays between relay-destination links. As long as source only
transmits in the first phase, different cooperative diversity
schemes differ only in the second phase on how relay nodes encode
 over that period. No matter which scheme is employed,
 the overall DM-tradeoff is always $3-6r$ provided the case of
 an empty set $\mathcal{D}(s)$ overshadows
other cases when more than one relay node succeeds in  decoding.
If this occurs, the overall DM-tradeoff is not affected by asynchronism.
 
\subsection{Distributed Delay Diversity}
\label{section: Delay Diversity}

In this section,  we first consider a scheme in which successful relay
nodes employ the same Gaussian codebook independent of
 the source codebook. We also investigate a repetition coding
based delay diversity scheme where relay nodes in $\mathcal{D}(s)$
use the same codebook adopted by source \cite{wei_ciss_04}.

It will be shown next in Theorem~\ref{theorem3} and
Theorem~\ref{theorem_delay} that as long as relative delay $T_0$
and transmitted signal bandwidth $B_w$ satisfies certain
conditions, both of these two schemes can achieve the same
DM-tradeoff as   the synchronous distributed space-time
coded scheme, which shows asynchronism does not hurt DM-tradeoff
in certain cases. In addition, we prove that repetition coding
based approach is fundamentally inferior than the independent
coding based approach due to its inefficiency in exploiting
degrees of freedom than  the former one, as revealed in
Theorem~\ref{theorem_delay}.

In \cite{winters_vtc_93}, a deliberate delay was also introduced between
two transmit antennas at a basestation in order to exploit the potential
spatial diversity.  Our proposed distributed delay diversity schemes
are similar with that scheme  in the sense both of these two approaches
create equivalent multipath link between transmitter and receiver. They
differ fundamentally, however, in the following ways: The relative
delays between transmit antennas at different relay nodes are inherent
in nature in our case due to distinct locations  of relay nodes, as
well as the difference in processing time  at each relay node. Secondly,
relative delays are required to satisfy certain conditions  in order to
achieve certain amount DM-tradeoff as proved in Theorem~\ref{theorem3}
and Theorem~\ref{theorem_delay}. These conditions imply higher layer
protocols should be implemented across relay nodes as proposed in
\cite{wei_jsac04}. While in \cite{winters_vtc_93} coordination through
protocols is not an issue as antennas are located at a basestation. The
last major difference is here  we are concerned with  the DM-tradeoff
function of diversity schemes. As a contrast, the diversity order
studied by \cite{winters_vtc_93} is only one particular point on the
DM-tradeoff curve for $r=0$. Therefore,  we term our schemes as {\em
distributed} delay diversity schemes in the sequel to avoid making any
further confusion.

\subsubsection{Independent Coding Based Distributed Delay Diversity}
\label{sec: indep. delay diversity}

In the system model described in  Section \ref{section: system model},
we assume   $\tau_1 \neq \tau_2$ and  $X_{R_1}(t) = X_{R_2}(t)
= X_{R}(t)$. Information bearing baseband signals $X_S(t) $
and $X_{R}(t)$, $t \in [0,T]$, are finite duration replica of two
independent stationary complex Gaussian random processes having zero
mean and  independent real and imaginary parts.  Their power spectral
densities (PSD) have  double-sided bandwidth $B_w/2$  and are assumed
to be flat since transmitters don't have side information about the
channel state and therefore `water-pouring'   \cite{gallagerbook}
cannot be used \cite{ozarow:1994}. Hence, the transmission of $X_j(t)$
equivalently leads to the transmission of $L= \lfloor B_w T \rfloor $
independent complex  Gaussian symbols over one packet \cite{ozarow:1994}
during each phase.  If there are more than one relay node in the decoding
set $\mathcal{D}(s)$, an equivalent multipath fading channel is formed
between these successful relay nodes and the destination in the second
phase.

When $ B_w T \gg 1$,  the  mutual information of the whole link given the
decoding set ${\mathcal D}(s)$, is 
\begin{gather} \label{eq: conditional mutual
infor.delay diversity non_repetition} I_{TDA}  =   \frac{1}{2} \log\left[ 1
+ \rho_0 |\alpha_{S,D}|^2 \right] + \nonumber \\ 
 \frac{1}{2B_w} \int_{-B_w/2}^{B_w/2}
\log\left[ 1 + \rho_0 | H_{R,D}(f) |^2 \right]\, df \, \mbox{(bits/s/Hz)
}, \end{gather}

\no where the second term is the mutual information of the
equivalent multipath fading channel whose frequency response is
$H_{R,D}(f) = \sum_{R_k \in {\mathcal D}(s)} \alpha_{R_k,D} e^{j 2
\pi f \tau_k} $  \cite{ozarow:1994} conditioned on fading gains
$\alpha_{i,j}$
and time delays $\{ \tau_k , R_k \in {\mathcal D}(s) \}$, and
$\rho_0 = \frac{2}{K+1} \mbox{SNR}$ is the normalized
signal-to-noise-ratio.

Given  delays $\{ \tau_k \}$,  the conditional outage probability is
 \be \label{eq: cond. outage prob. TDA given
tau} \mbox{P}_{out|\underline{\tau}} & = & \mbox{Pr} \left( I_{TDA} <
R | \underline\tau \right) \nonumber \\ 
& = & \sum_{{\mathcal D}(s)}
\mbox{Pr}\left[ {\mathcal D}(s)  | {\underline\tau} \right]
\mbox{Pr}\left[ I_{TDA} < R | {\mathcal D}(s),
\underline\tau\right], \ee

\no where  $R$ is defined in (\ref{eq: rate R}) and
$\underline\tau$ is the delay vector.
The outage probability averaged
over the distribution of delays  is \beeq \label{eq: definition
of final outage prob for TDA} \mbox{P}_{out} =   \mbox{Pr} \left(
I_{TDA} < R \right) = E_{\underline\tau}
\left[\mbox{P}_{out|\underline{\tau}}\right]. \eeeq

\no
Next, we show the asymptotic behavior of
$\mbox{P}_{out|\underline{\tau}}$ as $\mbox{SNR} \rightarrow \infty$
is   irrelevant of the exact values of
delays,  provided $\{ \tau_k \}$ satisfies certain conditions.

If the number of relay nodes forwarding in the second phase  is no
greater than $1$, i.e. $|\mathcal{D}(s)|\leq 1$, there does not exist
an equivalent multipath channel in the second phase and thus the mutual
information  $I_{TDA}$ in (\ref{eq: conditional mutual infor.delay
diversity non_repetition}) is equal to $I_{stc}$  determined in
(\ref{eq: STC mutual infor}) for the same decoding set ${\mathcal
D}(s)$. Therefore, the sum terms in (\ref{eq: cond. outage prob.
TDA given tau}) corresponding to $|{\mathcal D}(s)|=0$ and $|{\mathcal
D}(s)|=1$ have the same asymptotic slopes of $\mbox{SNR}$ as characterized
in (\ref{eq: whole link without relay}) and (\ref{eq: general link
outage with one realy node}). However, when two relay nodes are both
in $\mathcal{D}(s)$,  the mutual information $I_{TDA}$ in (\ref{eq:
conditional mutual infor.delay diversity non_repetition}) needs to be
studied individually.  Assume $\tau_k$ is put in an increasing order
and WLOG let $\tau_1 = \min_{R_k \in {\mathcal D}(s)} \tau_k =0$. Define
$T_0 = \min_{R_k \in {\mathcal D}(s), \tau_k \neq 0} \tau_k$.  We have

\begin{theorem}\label{theorem2}

As long as the relative delay between two paths $N_{R_1}
\rightarrow N_D$ and $N_{R_2} \rightarrow N_D$  satisfies $T_0 B_w
> 2$ and $T_0 B_w \notin \mathcal{Z}^+$, the conditional outage probability given $|\mathcal{D}(s)|=2$
satisfies
 \begin{gather} \label{eq: theorem2}
 \frac{1}{2} \prod_{k \in \left\{ S,R_1,R_2 \right\}} \tilde{\lambda}_{k,D}
\left( \widetilde{\mbox{SNR}} \right)^{-3 +  4r}
\stackrel{<}{\sim} \mbox{Pr} \left[ I_{TDA} < R, \, |{\mathcal
D}(s)| =2 | \underline{\tau} \right]  \nonumber \\ 
\stackrel{<}{\sim}
 2 \prod_{k \in \left\{ S,R_1,R_2 \right\}} \tilde{\lambda}_{k,D}
\left( \widetilde{\mbox{SNR}} \right)^{-3 +  6 r},  \end{gather}

\no for $r \in [0,1/2]$. If relative delay satisfies $T_0 B_w \in
\mathcal{Z}^+$, $\mbox{Pr} \left[ I_{TDA} < R, \, |{\mathcal
D}(s)| =2 | \underline{\tau} \right]$ vanishes  at a rate of
$\left( \widetilde{\mbox{SNR}} \right)^{-3 +  4 r}$ for large
$\mbox{SNR}$.
\end{theorem}

\begin{proof}

\no  Given $|\mathcal{D}(s)|=2$,  $I_{TDA}$ in (\ref{eq: conditional mutual infor.delay diversity
non_repetition}) can be expressed by 
\begin{gather} \label{eq: conditional
mutual infor. modified for TDA} I_{TDA} = \frac{1}{2} \log\left[ 1
+ \rho_0 |\alpha_{S,D}|^2 \right] +  \nonumber \\  \frac{1}{4 \pi B_w  T_0}
\int_{-\pi B_w T_0}^{\pi B_w T_0} \log\left[ 1 +  \rho_0 \left|
\sum_{k \in {\mathcal D}(s)} \alpha_{R_k, D} e^{j u
\frac{\tau_k}{T_0}} \right|^2 \right]\, du. \end{gather}

\no Note by Cauchy-Schwartz inequality, we have
 $\left|
\sum_{k \in {\mathcal D}(s)} \alpha_{R_k, D} e^{j u
\frac{\tau_k}{T_0}} \right|^2 \leq |{\mathcal D}(s)| \sum_{k \in
{\mathcal D}(s) } \left| \alpha_{R_k, D} \right|^2 $.  As a result,
the mutual information $I_{TDA}$ in (\ref{eq:
conditional mutual infor. modified for TDA}) can be
upper-bounded by $I^{(U)}_{TDA}$ defined below:
 \be
\label{eq: upperbound of the mutual infor. TDA} I_{TDA} & \leq &
\frac{1}{2}\log \left[ 1 + 2 \rho_0 \left( |\alpha_{R_1,D}|^2 +
|\alpha_{R_2,D}|^2\right) \right] +  \nonumber \\
& & 
\frac{1}{2} \log\left[ 1 + \rho_0 |\alpha_{S,D}|^2 \right] 
 \stackrel{\Delta}{=}  I^{(U)}_{TDA}.
 \ee

\no Comparing $I^{(U)}_{TDA}$
 with $I_{stc}$ in (\ref{eq: STC mutual infor}), we can see
$I^{(U)}_{TDA}$ is actually the mutual information of
 a synchronous space-time-coded cooperative diversity scheme with
$|\mathcal{D}(s)|=2$
and power scaled in the second phase. Therefore, the outage probability in
(\ref{eq: theorem2}) can be characterized by
 Lemma~\ref{lemma3}:
\beeq \label{eq: asymptotic of upper
bound of TDA} \mbox{Pr} \left[ I^{(U)}_{TDA} < R, \, |{\mathcal
D}(s)| =2, | \tau  \right] \sim
 \frac{1}{2} \prod_{k \in \left\{ S,R_1,R_2 \right\}} \tilde{\lambda}_{k,D}
\left( \widetilde{\mbox{SNR}} \right)^{-3 + 4 r}, \eeeq 

\no for $r \in [0,1/2)$,  which implies the
DM-tradeoff of the independent coding based distributed delay diversity  scheme
given $|\mathcal{D}(s)|=2$  cannot beat  the corresponding synchronous
space-time-coded approach, as expected.

Next, we seek a lower-bound of $I_{TDA}$.
Assume $T_0 B_w \geq 1$ and denote $\Delta_1 = \left\lfloor T_0
B_w \right\rfloor/ \lceil T_0 B_w \rceil \leq 1$, where $\left\lfloor x
\right\rfloor$ is the greatest integer less than or equal to  $x$ and
$ \lceil x \rceil $ is the smallest integer greater than or equal to $x$.
  The lower bound $I^{(L)}_{TDA} $of $I_{TDA}$
can be
determined as
\begin{gather} \label{eq: lowerbound of the mutual
infor. TDA} I_{TDA}  \geq \frac{\Delta_1}{2} \frac{1}{2 \pi
\delta_1} \int_{-\pi \delta_1}^{\pi \delta_1} \log\left[ 1 +
\rho_0 \left|  \sum_{k \in {\mathcal D}(s)} \alpha_{R_k, D} e^{j u
\frac{\tau_k}{T_0}}
\right|^2 \right]\, du \nonumber \\
+  \frac{1}{2} \log\left[ 1 + \rho_0
|\alpha_{S,D}|^2 \right] \nonumber \\
=   \frac{\Delta_1}{2}\log \left[\frac{1+ \rho_0 \nu + \sqrt{1+
\left( \rho_0 \omega \right)^2 + 2 \rho_0 \nu }}{2} \right]
\nonumber \\
+ \frac{1}{2} \log\left[ 1 + \rho_0 |\alpha_{S,D}|^2 \right] \nonumber \\
 \geq  \frac{\Delta_1}{2} \left( \log \left[ \frac{1 + \rho_0 \left(
|\alpha_{R_1,D}|^2 + |\alpha_{R_2,D}|^2\right)}{2} \right] \right.
\nonumber  \\
+ \left. \log\left[ 1 + \rho_0 |\alpha_{S,D}|^2 \right]  \right)
\stackrel{\Delta}{=} I^{(L)}_{TDA}, \end{gather}

\no where $\delta_1 = \left\lfloor T_0 B_w \right\rfloor$, $\nu = |\alpha_{R_1,D}|^2 +
|\alpha_{R_2,D}|^2 $ and $\omega =  |\alpha_{R_1,D}|^2 -
|\alpha_{R_2,D}|^2 $.  The  first inequality is due to the non-negative
 integrand in (\ref{eq: conditional
mutual infor. modified for TDA})  and $\Delta_1 \leq 1$.
 The equality  is
from the following integral equation \cite[pp. 527 (Eq. 41)]{table_integrals},
\beeq \label{eq: table integral}
\frac{1}{2\pi} \int_{0}^{2\pi} \log\left( 1 + a \sin x + b \cos x \right)\, dx
= \log \frac{1+ \sqrt{1-a^2 - b^2}}{2},\eeeq

\no for   $a^2 + b^2 <1$.
The last inequality is due to  $1 + \left( \rho_0 \omega \right)^2 + 2
\rho_0
\nu  \geq  0$ and $\Delta_1 \leq 1$.
Similar techniques used in proving Lemma~\ref{lemma3} can be applied to yield
\begin{gather} \label{eq: asymptotic of lower bound of TDA}
\mbox{Pr} \left[ I^{(L)}_{TDA} < R, \, |{\mathcal D}(s)| =2  |
\tau \right] \sim \nonumber \\
 2 \prod_{k \in \left\{ S,R_1,R_2 \right\}} \tilde{\lambda}_{k,D}
\left( \widetilde{\mbox{SNR}} \right)^{-3 + 4 \frac{r}{\Delta_1}}.
\end{gather}

\no If  $T_0 B_w$ is a positive integer,  we have $\Delta_1=1$ which
makes the lower bound and upper bound  of the outage probability
as shown  in  (\ref{eq: asymptotic of upper bound of TDA}) and
(\ref{eq: asymptotic of lower bound of TDA}) have the same asymptotic behavior.
If $T_0 B_w$ is a non-integer and $T_0 B_w > 2$, i.e. the relative
delay between two relay-destination links
satisfies $T_0> 2/B_w$, we have  $\Delta_1 \geq 2/3$
yielding
$3-4r/\Delta_1 \geq 3 - 6r$. Combining
(\ref{eq: asymptotic of upper bound of TDA}) and (\ref{eq:
asymptotic of lower bound of TDA}), therefore, yields Theorem~\ref{theorem2}.

\end{proof}

 Theorem~\ref{theorem2} essentially illustrates when two relay nodes both
succeed in decoding the source information and then forward it using the same
 Gaussian codebook  independent of what source sends, the overall
diversity gain is at least as good as $3-6r$ as long as the
relative delay $T_0$ between two paths is sufficiently large
satisfying the lower bound  $T_0 > 2/B_w$. This inequality reveals
a fundamental relationship featuring the dependence of performance
in terms of DM-tradeoff on the equivalent channel
characterizations.

If this condition on relative delay  is violated,  we are unable
to achieve the amount of diversity promised in
Theorem~\ref{theorem2}. For example, when $\tau_1=\tau_2$ i.e.
$T_0=0$,  signals transmitted by relay $1$ and $2$ will be
superposed at the destination end like a one-node relay channel
whose channel fading coefficient is $\alpha_{R_1,D} +
\alpha_{R_2,D}$. The resulting
 conditional outage probability $\mbox{Pr}
\left[ I_{TDA} < R, \, |{\mathcal D}(s)| =2  |
T_0=0 \right]$ thus has the same asymptotic relation as the one
with $|\mathcal{D}(s)|=1$ characterized by
Lemma~\ref{lemma1}, which implies the overall  diversity order is now dominated
by $2-2r$ and therefore  demonstrates the necessity and importance  of
satisfying the condition of $T_0 B_w \geq 2$. Since relative positions of nodes
do not necessarily ensure
 $T_0 B_w \geq 2$, a MAC layer protocol is required to   meet this requirement
\cite{wei_jsac04}.

Another remarkable point  is that the   condition in Theorem~\ref{theorem2} only
involves the relative delay $T_0$ and signal bandwidth $B_w$. This is because
 in our model we consider transmitting a bandlimited Gaussian random process
in a  continuous waveform channel and assume $B_w T \gg 1$ in order to invoke
the asymptotic results to obtain the closed form expression in (\ref{eq:
conditional mutual infor.delay diversity
non_repetition}) \cite{ozarow:1994, gallagerbook}. When transmitted signals
take the form of  linearly modulated cyclostationary random process as a 
practical  communication system does, the overall DM-tradeoff of delay
diversity will be addressed in Section~\ref{linearly modulated Delay-D} and
stated in Theorem~\ref{linearly modulated TDA}.

Given the  asymptotic behavior of
$\mbox{Pr} \left[
I_{TDA} < R, {\mathcal D}(s) \, | \, \underline{\tau} \right]$ for different
$|\mathcal{D}(s)|$, we are ready to calculate the 
overall DM-tradeoff. 

\begin{theorem}
\label{theorem3}
Given $T_0 B_w \geq 2$, where $T_0$ is the relative delay between two paths
from relay nodes to node $N_D$ and $B_w$ is the transmitted signal bandwidth,
 the DM-tradeoff of the distributed independent coding based delay diversity
scheme  is
 \be \label{eq: total asymptotic outage prob. on tau for TDA}
 d_{TDA}(r)&  = & 
\lim_{
\mbox{SNR} \rightarrow \infty} - \frac{ \log \left( \mbox{Pr}
\left[ I_{TDA} < R \left( \mbox{SNR} \right) \right] \right)}{\log
\mbox{SNR}}\nonumber \\
 & = &  3 ( 1 - 2 r) = d_{stc}(r), \, 0 \leq r < 1/2.
\ee
\end{theorem}

\begin{proof}

 When $|{\mathcal D}(s)|
\leq 1$, the rates of  this conditional outage probability
 decreasing to zero for large
$\widetilde{\mbox{SNR}}$ are equal to those for the  corresponding
distributed synchronous space-time-coded scheme, i.e. diminishing
rates of  $\mbox{Pr} \left[ I_{TDA} < R, {\mathcal D}(s) \, | \,
\tau \right]$ are in the order of 
  $\left( \widetilde{\mbox{SNR}} \right)^{-3 + 6r}$ and
  $\left( \widetilde{\mbox{SNR}}\right)^{-3 + 4 r}$ for $|{\mathcal
D}(s)|=0$ and $|{\mathcal D}(s)| =1$, respectively. When
$|\mathcal{D}(s)|=2$, as long as $T_0 B_w \geq 2$, $\mbox{Pr}
\left[ I_{TDA} < R, {\mathcal D}(s) \, | \, \underline{\tau} \right]$
decreases to zero at least in the order of  $\left( \widetilde{\mbox{SNR}} \right)^{-3+6r}$
from
Theorem~\ref{theorem2}. Therefore, as far as the overall
DM-tradeoff is concerned, $3-6r$
 is the dominant term
determining the  slope of the total outage probability
$P_{out|\underline{\tau}}$ in (\ref{eq: cond. outage prob. TDA given tau})
decreasing to  zero given  $T_0 B_w \geq 2$.

 Moreover, we can see if  $T_0 B_w \geq 2$,
 bounds in Theorem~\ref{theorem2}
do not depend on the exact value of $T_0$,  which implies $E_{\underline{\tau}}
\left[P_{out|\underline{\tau}} \right]$ in (\ref{eq: definition
of final outage prob for TDA}) has the same asymptotic
dominant term
$\widetilde{\mbox{SNR}}^{-(3-6r)}$.  Therefore, even at the presence of non-zero
relative delays,  the same DM-tradeoff  as
the synchronized space-time-coded cooperative diversity scheme  can still be
achieved, which proves Theorem~\ref{theorem3}.

\end{proof}

Note we restrict ourselves to the case of having only two relay
nodes. For cases having more than two relay nodes, the analysis
will be more involved and we expect there will exist a lower bound
on the minimum relative delay among multipath from each relay node
to the destination in order to yield a satisfying DM-tradeoff
lower bound.
\subsubsection{Repetition Coding Based Distributed Delay Diversity}
\label{sec: repetition delay diversity}

For the purpose of simplicity,  relay nodes in the decoding set can also
use the same codeword employed by source instead of using an independent
codebook. In this section, we look into the DM-tradeoff
of such a repetition coding based
distributed delay diversity approach. 

 Denote $I_{R-TDA}$ as the mutual information of
this relay channel. It can be shown
\cite{ozarow:1994}
\beeq \label{eq: conditional
mutual infor.delay diversity repetition} I_{R-TDA} =
\frac{1}{2B_w} \int_{-\frac{B_w}{2}}^{\frac{B_w}{2}} \log\left[ 1 +  \rho_0
|\alpha_{S,D}|^2 + \rho_0 | H_{R,D}(f) |^2 \right]\, df, \eeeq

\no where $H_{R,D}(f)$ is  defined  after (\ref{eq: conditional
mutual infor.delay diversity non_repetition}).
Next, we investigate the asymptotic behavior of
 $\mbox{Pr}
\left[I_{R-TDA} < R, \right.$ $\left.  |{\mathcal D}(s)| =  j\, |
\, \underline{\tau}   \right]$, $j=0,1,2$.

For $|{\mathcal
D}(s)|=0$, we have $I_{R-TDA} = I_{stc} = \frac{1}{2}\log(1+ \rho_0 |\alpha_{S,D}|^2)$
whose outage probability has the same asymptotic characteristic
as in  (\ref{eq: whole link without relay}).
 When $|{\mathcal
D}(s)| =1$, we have $I_{R-TDA} = \frac{1}{2} \log \left[ 1  +
\rho_0 \left( |\alpha_{S,D}|^2 \right. \right.$  $\left. \left.+
|\alpha_{R_j,D}|^2 \right) \right]$, where $N_{R_j} \in {\mathcal
D}(s)$. The sum of two
 independently distributed exponential random variables $\left(|\alpha_{S,D}|^2
+ |\alpha_{R_j,D}|^2 \right)$ has the similar asymptotic pdf as
specified in (\ref{ eq: pdf of the sum of two exp. rv}). The outage probability
in this case is characterized by Lemma~\ref{lemma4}.

\begin{lemma} \label{lemma4}
When there is only one relay node in $\mathcal{D}(s)$, the asymptotic
equivalence of the outage probability for  repetition coding based distributed
delay diversity is
\begin{gather} \label{eq: asymptotic I_{R-TDA} for |D(s)| =1} \mbox{Pr}
\left[I_{R-TDA} < R, |{\mathcal D}(s)| = 1 \, | \, \underline{\tau}   \right]
\nonumber \\ 
\sim \left( \tilde{\lambda}_{S,R_1}\tilde{\lambda}_{R_2,D} +
\tilde{\lambda}_{S,R_2}\tilde{\lambda}_{R_1,D} \right)
\tilde{\lambda}_{S,D} \widetilde{\mbox{SNR}}^{-3(1-2r) }. \end{gather}
\end{lemma}

 \begin{proof}
Combining the asymptotic result on the decoding set probability in (\ref{eq:
asymptotic of decoding set}) for $|\mathcal{D}(s)|=1$
and slight modifying the proof of Lemma~\ref{lemma3},
we obtain
the RHS of (\ref{eq: asymptotic I_{R-TDA} for |D(s)| =1}).

\end{proof}

\no If both relay nodes are in  ${\mathcal D}(s)$, the repetition
coding based mutual information $I_{R-TDA}$ is
\cite{ozarow:1994}, \be \label{eq: conditional mutual infor.
modified for R-TDA} I_{R-TDA} & = & \frac{1}{4 \pi B_w  T_0}
\int_{-\pi B_w T_0}^{\pi B_w T_0} \log\left[ 1 +   \rho_0
|\alpha_{S,D}|^2  \right. \nonumber \\ & & \left. + \rho_0 \left|  \sum_{k \in {\mathcal D}(s)}
\alpha_{R_k, D} e^{j u \frac{\tau_k}{T_0}} \right|^2 \right]\, du.
\ee

\no Applying the bounding techniques developed  for $I_{TDA}$ when
$|{\mathcal D}(s)| =2$, we obtain   \be \label{eq: the upper and
lower bound of R-TDA with 2 relay nodes}
 Q_0\left(\tilde{\lambda}\right)
\left( \widetilde{\mbox{SNR}} \right)^{-3 +  6 r}
& \stackrel{<}{\sim} & \mbox{Pr} \left[ I_{R- TDA} < R, \, |{\mathcal
D}(s)| =2 | \underline{\tau} \right] \nonumber \\ 
& \stackrel{<}{\sim} & 
 Q_1 \left(\tilde{\lambda}\right)
\left( \widetilde{\mbox{SNR}} \right)^{-3 +  6 r/\Delta_1} , \ee

\no where  $Q_0 \left(\tilde{\lambda}\right) =\frac{1}{4}
\frac{2\tilde{\lambda}_{S,D} - \tilde{\lambda}_{R_1,D}}
{2\tilde{\lambda}_{S,D} - \tilde{\lambda}_{R_2,D}}  \prod_{k \in
\left\{ S,R_1,R_2 \right\}} \tilde{\lambda}_{k,D}$, $Q_1
\left(\tilde{\lambda}\right) =\frac{\tilde{\lambda}_{S,D} -
\tilde{\lambda}_{R_1,D}} {\tilde{\lambda}_{S,D} -
\tilde{\lambda}_{R_2,D}}  \prod_{k \in \left\{ S,R_1,R_2 \right\}}
\tilde{\lambda}_{k,D}$  given
$\tilde{\lambda}_{S,D} > \tilde{\lambda}_{R_1,D}
> \tilde{\lambda}_{S,D}$. For other situations regarding $\left\{
\tilde{\lambda}_{j,D} \right\}$,
 we have the similar lower- and upper-bounds as in
(\ref{eq: the upper and lower bound of R-TDA with 2 relay nodes})
except    functions $Q_0
\left(\tilde{\lambda}\right)$ and  $Q_1
\left(\tilde{\lambda}\right)$ need to be modified accordingly without affecting
slopes.

Based on the asymptotic equivalence of conditional outage probability
for cases $|{\mathcal D}(s)| =j, j=0,1,2$, as shown in  (\ref{eq: whole
link without relay}), (\ref{eq: asymptotic I_{R-TDA} for |D(s)| =1})
and (\ref{eq: the upper and lower bound of R-TDA with 2 relay nodes}),
respectively, we can conclude about the overall diversity gain for the
repetition coding based distributed delay diversity:

\begin{theorem}\label{theorem_delay}
The upper-bound and lower-bound of the overall DM-tradeoff of the
repetition coding based distributed delay diversity are determined
by

\beeq \label{eq: diversity gain asymptotic of I_{R-TDA}}
 3-6 r /\Delta_1 \leq  d_{R-TDA}(r)  \leq  3 - 6r = d_{TDA}(r) , \, 0 \leq  r <
\frac{1}{2},
\eeeq

\no  where $\Delta_1 = \left\lfloor T_0 B_w \right\rfloor/ \lceil T_0 B_w
\rceil \leq 1$, provided the relative delay $T_0$ and transmitted signal
bandwidth $B_w$ satisfies $T_0 B_w \geq 1$.  The equality in (\ref{eq:
diversity gain asymptotic of I_{R-TDA}}) is achieved when $T_0 B_w \in
{\mathcal Z^+}$, i.e. when $\Delta_1 =1$.  \end{theorem}

\begin{proof}

First, the lower bound in  (\ref{eq: the upper and lower bound of
R-TDA with 2 relay nodes}) demonstrates $\mbox{Pr} \left[ I_{R-
TDA} < R, \, |{\mathcal D}(s)| =2 | \underline{\tau} \right]$
decreases to zero  no faster than $\left( \widetilde{\mbox{SNR}}
\right)^{-3+6r}$, which is the vanishing rate for cases of
$|\mathcal{D}(s)| \leq 1$, as reflected in (\ref{eq: whole link
without relay}) and  (\ref{eq: asymptotic I_{R-TDA} for |D(s)|
=1}). We can thus infer that the dominant factor affecting the
overall DM-tradeoff is subject to the case of $|{\mathcal D}(s)|
=2$, which consequently yields the inequality in (\ref{eq:
diversity gain asymptotic of I_{R-TDA}}).

If the relative delay and transmitted signal bandwidth satisfies
$T_0 B_w \geq 1$, we have  $1 \geq  \Delta_1 \geq 1/2$; otherwise
$\Delta_1 =0$ making the lower bound in (\ref{eq: diversity gain
asymptotic of I_{R-TDA}}) trivial. Meanwhile,  when $T_0 B_w$ is a
positive integer, the asymptotic rates reflected in  the lower and
upper bounds in (\ref{eq: the upper and lower bound of R-TDA with
2 relay nodes}) agree with each other, which yields $ d_{R-TDA}(r)
= 3- 6r$.

We can therefore conclude
based upon the preceding analysis  that the diversity  of
repetition  coding based distributed delay diversity scheme is
always no greater than  the independent coding based distributed delay
diversity scheme, and thus complete proof of
Theorem~\ref{theorem_delay}.

\end{proof}

In terms of DM-tradeoff,
Theorem~\ref{theorem_delay} reveals a fundamental limitation imposed by
employing the {\em repetition} coding based  relaying strategy as compared with
the {\em independent } coding based one in Theorem~\ref{theorem3}.
An additional observation we can make from Theorem~\ref{theorem_delay}  and
Theorem~\ref{theorem3} is that distributed delay diversity schemes achieve the
same DM-tradeoff $3-6r$ as that under  synchronous distributed
space-time-coded cooperative diversity approach studied in\cite{laneman_03}, if
the relative delay $T_0$ and bandwidth $B_w$ satisfies $T_0 B_w \in
\mathcal{Z^+}$.  Moreover, if $T_0 B_w \geq 2$, both of these two cooperative
diversity schemes achieve a  diversity of order $3$, the
number of potential transmit nodes, when the spectral efficiency $R$ remains
fixed with respect to $\mbox{SNR}$, i.e. $r=0$, which further demonstrates
asynchronism does not hurt diversity as long as the relative delay is sufficiently
big to allow us to exploit spatial diversity.

\subsubsection{Distributed Delay Diversity with Linearly Modulated Waveforms}
\label{linearly modulated Delay-D}

For the distributed delay diversity schemes analyzed in Section
\ref{sec: indep. delay diversity} and \ref{sec: repetition delay diversity}, 
the transmitted information
carrying  signal $X_j(t)$ is assumed to be a finite duration
replica of a complex stationary Gaussian random process
 with a flat power spectral
density, which is widely adopted  in studying the capacity of frequency
selective fading channel \cite{ozarow:1994}. In this section, we study
 the diversity gain of an independent coding based  distributed delay
diversity scheme employing  linearly modulation waveforms
 $X_j(t)=
\sum_{k =1}^{n}
b_{j}(k) s_j(t - kT_s)$ for  $j \in \{
S, R_1, R_2\}$, where $s_j(t)$ is a strictly time limited
and  root mean squared (RMS) bandlimited waveform  \cite{roger_91} of
duration $T_s$,
with unit energy $\int_0^{T_s} |s_j(t)|^2 \, dt  =1$ ( $T_s$ is the symbol
period), and  $b_j(k)$ is the $k$th symbol transmitted by the $j$th
user satisfying the following power constraint: $\frac{1}{n}
\sum_{k=1}^{n} b_j^2(k) \leq p_j$, with $p_j = \frac{2}{K+1}
\hat{P}_s$. This linearly modulated waveform model is often employed to study
the capacity of  asynchronous multiuser systems
\cite{verdu_89_mac,roger_91,roger_92} and will be adopted as well when we
investigate the  DM-tradeoff of our proposed asynchronous space-time coded
cooperative diversity scheme in Section~\ref{section: asyn. coding}.

Assume relay nodes employ the decode-and-forward strategy under which  $\{
b_{R_1}(k) = b_{R_2}(k) \}$ is a sequence of  i.i.d complex
Gaussian random variables with zero mean and unit variance, and
independent of $\{ b_{S}(k) \}$. Let $I_{L-TDA}$ denote the
 mutual information of an entire link, which can be computed as in
 (\ref{eq: conditional mutual infor.delay diversity non_repetition}): 
\beeq I_{L-TDA} = \frac{1}{2}
\log \left( 1 + \rho_0 |\alpha_{S,D}|^2 \right) + \frac{1}{2}
I_{2-TDA} \eeeq

\no where $I_{2-TDA}$ is defined as the mutual information of the
equivalent channel between two relays and destination node.

When there is no more than one relay  node involved in forwarding,   the outage
probability  $P\{I_{L-TDA} < R, \mathcal{D}(s) \}$ for $|
\mathcal{D}(s)| \leq 1$ has the same asymptotic
behavior as $P\{I_{TDA} < R, \mathcal{D}(s)\}$ obtained in
Section~\ref{sec: indep. delay diversity} in same cases.
We thus  focus only  on the case of $|\mathcal{D}(s)|=2$. 

\begin{theorem} \label{linearly modulated TDA}
For the independent coding based distributed delay diversity
scheme under a relative delay
$\tau \in (0, T_s]$,  if $X_j(t)$ is linearly modulated using a time-limited
waveform $s(t)$ of duration $T_s$ , the outage probability ${\mbox Pr}\left[
I_{L-TDA}<R, |\mathcal{D}(s)|=2, \right]$ has the following
asymptotic equivalence,
\begin{gather} \label{eq: linearly modulated TDA}
\mbox{Pr} \left[ I_{L-TDA} < R, \; |{\mathcal D}(s)| =2 \right] \nonumber \\
\sim   \frac{2}{\sqrt{1- |\rho_{12}|^2}}  \prod_{k \in \left\{ S,R_1,R_2 \right\}} \tilde{\lambda}_{k,D}
\left( \widetilde{\mbox{SNR}} \right)^{-3 + 4 r},  \end{gather}

\no for $ 0 \leq r \leq 1/2$, where $|\rho_{12}|= |\int_0^{T_s} s(t) s(t-\tau) \, dt| <1$  and $
\tilde{\lambda}_{k,D}$ are defined in
Section~\ref{section: syn. distc}.
\end{theorem}

\begin{proof}
The proof is given  in  Appendix~\ref{theoremLTDAProof}.
\end{proof}

 Theorem~\ref{linearly modulated TDA}  demonstrates when
$X_j(t)$ is linearly modulated using $s(t)$ of duration  $T_s$ and two relay
nodes are both in $\mathcal{D}(s)$, the independent coding based distributed
delay diversity scheme achieves a  diversity of order $3-4r$. This result shows
under certain conditions  asynchronism does not
affect the DM-tradeoff when  compared with the synchronous space-time
coded approach as revealed in Lemma~\ref{lemma3}. When we count
all possible outcomes of the relay decoding to calculate the
overall DM-tradeoff function,  we obtain
$d_{L-TDA}(r)=3-6r = d_{stc}(r), \, r \in [0, 1/2]$ due to the same dominating
factor caused by no relay nodes forwarding source information as
observed in previous sections.

\subsection{Asynchronous Space-time-coded Cooperative Diversity}
\label{section: asyn. coding}

In this section, assuming no synchronization among relay nodes, we
propose a more spectral efficient approach termed as {\em
asynchronous space-time-coded diversity scheme} to exploit the
spatial diversity in relay channels. This approach has a better
DM-tradeoff than both the distributed delay  diversity and
synchronous space-time-coded schemes when two relay nodes are both
in the decoding set ${\mathcal D}(s)$. Actually, we will show
under certain conditions on the baseband waveform used by both
relay nodes, the link between source and its destination across
two relay nodes is equivalent to a parallel channel consisting of
three independent channels in terms of the overall DM-tradeoff
function. As a result,
 employing asynchronous space-time codes enables us
to fully
exploit all degrees of freedom
available in the {\em space-time} domain in relay channels.

We  divide the major proof into $3$ steps to streamline our
presentation. First, we set up an equivalent discrete time
channel model from which we obtain the sufficient statistics for
decoding under symbol level asynchronism. Next, we prove a
convergence result for the achievable  mutual information rate as
the codeword block length goes to infinity by applying some
techniques in asymptotic spectrum distribution of  Toeplitz forms.
Finally, we prove a sufficient condition for the existence of a
strictly positive minimum eigenvalue of the Toeplitz form involved
in the former asymptotic mutual information rate. The existence of
such positive minimum eigenvalue proves to be crucial in showing
an equivalence of the relay-destination link to a parallel channel
consisting of two independent users, and thus leads us to the
desired result on DM-tradeoff function. At the end of this
section, we will make remarks on some cases where not only does
asynchronous coded approach perform better than  synchronous one in terms of
DM-tradeoff, but also it results in strictly greater capacity than
synchronous one when both relay nodes succeed in decoding.

\subsubsection{Discrete Time System Model for Asynchronous Space-time Coded
Approach}

To address  the impact of asynchronism,
we follow the footsteps of \cite{verdu_89_mac} by assuming a time-limited
baseband waveform. What distinguishes us from \cite{verdu_89_mac} is our
approaches and results are valid for time constrained
waveforms of an arbitrary finite duration, while \cite{verdu_89_mac} requires a waveform lasting for one symbol period.
To gain insights and WLOG, we  first tackle a problem where the baseband
waveforms employed are time-limited
 within $2$ symbol periods, and then extend the
results to the case  with any arbitrarily  time-limited waveforms. The
transmitted baseband signals are $X_j(t)= \sum_{k =1}^{n} b_{j}(k)
s_j(t - kT_s)$, $j \in \left\{ S, R_1, R_2\right\}$ where $s_j(t)$
is a time-limited waveform of duration $2T_s$ with unit energy,
i.e. $\int_0^{2T_s} |s_j(t)|^2 \, dt  =1$,  and $b_j(k)$ is the $k$th symbol
transmitted by the $j$th
user satisfying the same  power constraint described in Section~\ref{linearly
modulated Delay-D}.
We assume $X_j(t)$ lasts over a duration of length $T$ and
the number of symbols transmitted  $n= T/T_s$ is
sufficiently large, i.e. $n \gg 1$, such that the later mutual
information has a convergent closed form.

When two relay nodes both succeed in decoding the source messages,
asynchronous space-time-codes are encoded across them to forward the
source messages to the destination. Without any channel state information
of the link between $N_{R_j}$ and $N_D$, independent i.i.d complex
Gaussian codebooks are assumed which are independent of the source
codebook. The main difference from the traditional space-time codes is
the asynchronous one encodes without requiring signals arriving at the
destination from virtual antennas (i.e. relay nodes) to be perfectly
synchronized

Let $I_{A-stc}$ denote the mutual information of the source-destination channel
under the proposed asynchronous space-time-coded scheme.
The
outage probability of the whole link is
\beeq \label{eq: total outage for Asyn. st. coding}
\mbox{Pr}\left[ I_{A-stc} < R \right] = \sum_{j = 0}^2
\mbox{Pr}\left[ I_{A-stc} < R, \, |{\mathcal D}(s)| = j \right].
\eeeq

\no As only when $|{\mathcal D}(s)|=2$ will we  consider  the
issue of encoding across relay nodes and cases of $|{\mathcal
D}(s)| \leq 1$ are identical as the corresponding cases for
synchronous space-time coded approach, we first focus on the case
of $|\mathcal{D}(s)|=2$.
Given $|\mathcal{D}(s)|=2$,  we obtain
\beeq  \label{eq: the first whole link
mutual infor. for asyn.s-time codes} I_{A-stc}
=  \frac{1}{2} I_{E-SD}
+ \frac{1}{2} I_{E-MacA}, \eeeq

\no where  $I_{E-SD}$ is the mutual information of the direct link
channel when the baseband waveform has  finite duration, and $I_{E-MacA}$
is the mutual information of a $2 \times 1$ MISO system featuring
the communication link  between  two successful relay nodes and the
destination at the presence of symbol level asynchronism caused  by the
relative delay $\tau_2 - \tau_1$, which is assumed to satisfy $T_s >
\tau_2 - \tau_1 > 0$. If the relative delay is greater than $T_s$,
this does not affect $I_{E-MacA}$  for asymptotically long codeword
\cite{verdu_89_memory}.  Our objective  is to study the asymptotic
behavior of $I_{E-MacA}$ for large $n$ since this is closely related
to the asymptotic analysis of outage provability conditioned on
$|\mathcal{D}(s)|=2$.

Next, we develop an equivalent discrete time system model.
Assuming  $\tau_j$ are known to the destination perfectly,  we
obtain sufficient statistics for making decisions on transmitted
data vector $\{b_1(k),  b_2(k)\}, k=1, \cdots, n$ by passing the
received signals through two matched filters for signals
$s_j(t-\tau_j)$, respectively \cite{verdu_89_mac}.  The sampled matched filter outputs
are
\beeq \label{eq: sufficient stat. for extended asyn st.coding}
y_{D_{R_j}} (k) = \int_{k T_s + \tau_j}^{(k+2)T_s + \tau_j}
y_{D_R}(t) \alpha^{*}_{R_j,D} s_j\left( t-kT_s-\tau_j \right) \,
dt, \eeeq

\no for $j =1,2, \; k= 1, \cdots, n.$ 

\no  Given $T_s > \tau_2 -\tau_1 > 0$, the equivalent
discrete-time system model is characterized by
\begin{gather} \label{eq: discrete time extended asynchro. model} \left[
\begin{array}{l}
y_{D_{R_1}}(k) \\
y_{D_{R_2}}(k)
\end{array}
\right]  =  \left[ \begin{array}{cc}
0 & c_2 \alpha^{*}_{R_1} \alpha_{R_2}  \\
0 & 0
\end{array}
\right]  \left[ \begin{array}{c}  b_{R_1} (k -2)  \\ b_{R_2}(k-2)
\end{array} \right] \nonumber \\
+
\left[ \begin{array}{cc}
a_1 |\alpha_{R_1}|^2  & c_1 \alpha^{*}_{R_1} \alpha_{R_2}   \\
 f_1 \alpha^{*}_{R_2} \alpha_{R_1}  &  |\alpha_{R_2}|^2 d_1
\end{array}
\right]  \left[ \begin{array}{c}  b_{R_1} (k-1)  \\ b_{R_2}(k-1)
\end{array} \right]
\nonumber \\
 + \left[ \begin{array}{cc}
|\alpha_{R_1}|^2  & c_0 \alpha^{*}_{R_1} \alpha_{R_2}   \\
 c_0 \alpha^{*}_{R_2} \alpha_{R_1}  &  |\alpha_{R_2}|^2
\end{array}
\right]  \left[ \begin{array}{c}  b_{R_1} (k)  \\ b_{R_2}(k)
\end{array} \right] \nonumber \\
 + \left[ \begin{array}{cc}
a_1 |\alpha_{R_1}|^2  & f_1 \alpha^{*}_{R_1} \alpha_{R_2}   \\
c_1 \alpha^{*}_{R_2} \alpha_{R_1}  &  |\alpha_{R_2}|^2 d_1
\end{array}
\right]  \left[ \begin{array}{c}  b_{R_1} (k+1)  \\ b_{R_2}(k+1)
\end{array} \right]
\nonumber \\
  + \left[ \begin{array}{cc}
0 & 0 \\
c_2  \alpha^{*}_{R_2} \alpha_{R_1}  &  0
\end{array}
\right]  \left[ \begin{array}{c}  b_{R_1} (k+2)  \\ b_{R_2}(k+2)
\end{array} \right] + \left[ \begin{array}{c}  z_{R_1} (k)  \\
z_{R_2} (k) \end{array} \right],  \end{gather}

\no for $\, k =1, \cdots, n,$ 
 with $b_{R_j}(0)=b_{R_j}(-1)=b_{R_j}(n+1)=b_{R_j}(n+2)=0,
j=1,2$. The coefficients of $c_1$, $a_1$, $f_1$ and $d_1$
are defined as
\begin{gather}\label{eq: coefficients a_1 and d_1} a_1 =
\int_0^{T_s} s_1(t) s_1(t + T_s), \; d_1 = \int_0^{T_s} s_2(t) s_2(t
+ T_s), \\
c_0 = \int_0^{2T_s} s_1(t) s_2\left(t - \tau_2 +\tau_1
\right), \\  c_1 = \int_0^{2T_s} s_1(t) s_2\left(t +T_s +\tau_1-
\tau_2  \right)\\
 f_1 = \int_0^{T_s} s_2(t) s_1
\left( t+ T_s + \tau_2 -\tau_1 \right),\label{eq: coeef. f_1 c_2} \\
c_2 = \int_0^{T_s}
s_1(t) s_2 \left( t+ 2T_s - \tau_2 + \tau_1 \right). \end{gather}

\no Thus, the original $2\times 1$ MISO channel is now transformed into
 a $2\times 2$ MIMO channel in the
discrete time domain with vector inter-symbol-interferences (ISI).
The additive noise vector    $\left[  z_{R_1} (k), \, z_{R_2} (k)
\right]^T$  in (\ref{eq: discrete time extended asynchro. model})
is a discrete time  Gaussian random process  with zero mean and
covariance matrix  \beeq \label{eq: covariance matrix of sampled noise
for asyn. st. coding} E \left[\left[ \begin{array}{c}  z_{R_1} (k)  \\
z_{R_2} (k) \end{array} \right] \left[ z_{R_1}^{*}(l),\; z_{R_2}^{*}(l)
\right] \right] = {\mathcal N}_0 {\bf H_E}(k-l), \eeeq

\no where  ${\bf H_E}(i)$ for  $|i| > 2$ are all zero matrices, and matrices
${\bf
H_E}(j),\, -2 \leq j \leq 2$ are  
\begin{gather} \label{eq: matrices H for extended Asynchronous St. coding}
{\bf H_E}(0) = \left[
\begin{array}{cc}
|\alpha_{R_1}|^2 & c_0 \alpha^{*}_{R_1} \alpha_{R_2}   \\
 c_0 \alpha^{*}_{R_2} \alpha_{R_1}  &  |\alpha_{R_2}|^2
\end{array}
\right], \\
 {\bf H_E}(1) = {\bf H_E}^{\dagger}(-1) = \left[
\begin{array}{cc}
a_1 |\alpha_{R_1}|^2  & c_1 \alpha^{*}_{R_1} \alpha_{R_2}   \\
 f_1 \alpha^{*}_{R_2} \alpha_{R_1}  &  |\alpha_{R_2}|^2 d_1
\end{array}
\right], \\
{\bf H_E}(2) = {\bf H_E}^{\dagger}(-2) =
\left[ \begin{array}{cc}
0  & c_2 \alpha^{*}_{R_1} \alpha_{R_2}   \\
0  & 0
\end{array}
\right], 
\label{eq: second part of matrices H for extended Asynchronous St. coding} 
\end{gather}

\no where $A^{\dagger}$ is the conjugate transpose of a matrix
$A$.

Denote $\underline{y}_{D_R}(k) = \left[ y_{D_{R_1}}(k),
y_{D_{R_2}}(k) \right]$, $\underline{b}_{R}(k) = \left[
b_{R_1}(k), b_{R_2}(k) \right]$ and $\underline{z}_R(k) = \left[
z_{R_1} (k), \,  \right.$ $\left. z_{R_2} (k) \right]$ for $k=1,
\cdots, n$.  The discrete time system model of (\ref{eq: discrete
time extended asynchro. model}) can be expressed in a more compact
form by
 \beeq \label{eq: compact form
of the discrete time extended asynchro. model} \underline{\bf y}^n
= {\bf \mathcal H_E}\underline{\bf b}^n + \underline{\bf z}^n,
\eeeq

\no where \beeq \underline{\bf y}^n =\left[
\underline{y}_{D_R}(1),\: \underline{y}_{D_R}(2),\:
\: \cdots,
 \underline{y}_{D_R}(n)
 \right]^T,
\eeeq

\no \beeq \underline{\bf b}^n =\left[ \underline{b}_{R}(1),\:
\underline{b}_{R}(2),\:
\cdots,  
 \underline{b}_{R}(n)
 \right]^T,
\eeeq

\no  \beeq \underline{\bf z}^n =\left[ \underline{z}_{R}(1),\:
\underline{z}_{R}(2),\: \cdots,  \underline{z}_{R}(n)
 \right]^T,
\eeeq

\no and  ${\bf \mathcal H_E} $ is a Hermitian block Toeplitz
matrix defined by
 \beeq
 {\bf \mathcal H_E}  =
\left[  \begin{array}{ccccc}
{\bf H_E}(0) & {\bf H_E}(-1) & {\bf H_E}(-2) & \multicolumn{2}{c}{} \\
{\bf H_E}(1) & {\bf H_E}(0) &  {\bf H_E}(-1) & {\bf H_E}(-2) &
\multicolumn{1}{c}{} \\
{\bf H_E}(2) & {\bf H_E}(1) & {\bf H_E}(0) &  {\bf H_E}(-1) & {\bf H_E}(-2) \\
\multicolumn{5}{c}{\dotfill} \\
\multicolumn{2}{c}{} &  {\bf H_E}(2) & {\bf H_E}(1) & {\bf H_E}(0) \\
\end{array}
\right], \eeeq

\no which is also the covariance matrix of the Gaussian vector
$\underline{\bf z}^n$.

Suppose ${\bf \mathcal H_E}^n$ is available only at the
destination end and  transmitters employ independent complex
Gaussian codebooks, i.e. vectors $\underline{\bf b}_{R_1} = \left[
b_{R_1}(1), \cdots, b_{R_2}(n) \right]^T$ and $\underline{\bf
b}_{R_2} = \left[ b_{R_1}(1), \cdots, b_{R_2}(n) \right]^T$ are
independently distributed proper complex white Gaussian vectors,
the mutual information of this equivalent $2 \times 2$ MIMO system
at the presence of memory introduced by ISI is \cite{gallagerbook}
 \be\label{eq: sum  mutual
information of the extended  MAC channel} I^{(n)}_{E-MacA} &  = &
\frac{1}{n} I \left(\underline{\bf y}^n ; \underline{\bf b}^n
\right)\nonumber \\ & = & \frac{1}{n}\log \mbox{det} \left [ {\bf I}_{2n} +
\frac{1}{{\mathcal N}_0} E \left[ \underline{\bf b}^n
\left(\underline{\bf b}^n\right)^{\dagger} \right] {\bf \mathcal
H_E} \right]. \ee

\subsubsection{Convergence of  $I^{(n)}_{E-MacA}$ as $n \rightarrow \infty$}

To obtain the asymptotic result  of $I^{(n)}_{E-MacA}$ as $n$
approaches infinity, we can rewrite the matrix ${\bf \mathcal H_E}$
as ${\bf \mathcal H_E}  =
 {\bf P}^n {\bf \mathcal T}^{(2n)} \left({\bf P}^n\right)^T$, where
${\bf \mathcal T}^{(2n)}$ is  a  Hermitian  block matrix
\cite{gazzah_01_block_toeplitz} defined by
\[ {\bf \mathcal T}^{(2n)} = \left[
\begin{array}{cc}
|\alpha_{R_1,D}|^2 {\bf T_E}^n (1,1) & \alpha_{R_1,D} \alpha^*_{R_2,D} {\bf T_E}^n (1,2)  \\
\alpha^{*}_{R_1,D} \alpha_{R_2,D}{\bf T_E}^n (2,1) &
|\alpha_{R_2,D}|^2{\bf T_E}^n (2,2)
\end{array}\right]
\]

\no  and ${\bf P}^n $ is a permutation matrix such that ${\bf P}^n
\underline{\bf b}^n$ is a column vector of dimension $2n$ whose
first
 and second half entries are $\underline{\bf b}_{R_1}$ and $\underline{\bf b}_{R_2}$,
respectively. The block matrices ${\bf T_E}^n (i,j),  i, j \in
\left\{1, 2\right\}$ are $n\times n$ Toeplitz matrices specified as
\begin{gather}
\label{eq: matrix T_n (1,1)} {\bf T_E}^n (1,1) = \left[\begin{array}{cccccc}
1 & a_1 & 0 & \multicolumn{3}{c}{} \\
a_1 & 1 & a_1 & 0 & \multicolumn{2}{c}{}  \\
0 & a_1 & 1 & a_1 & 0 &   \\
\multicolumn{6}{c}{\dotfill} \\
\multicolumn{4}{c}{}  & a_1 & 1\\
\end{array}
\right], \nonumber \\
{\bf T_E}^n (2,2) =  \left[\begin{array}{cccccc}
1 & d_1 & 0 & \multicolumn{3}{c}{} \\
d_1 & 1 & d_1 & 0 & \multicolumn{2}{c}{}  \\
0 & d_1 & 1 & d_1 & 0 &   \\
\multicolumn{6}{c}{\dotfill} \\
\multicolumn{4}{c}{}  & d_1 & 1\\
\end{array}
\right] \end{gather}

\no and \beeq \label{eq: matrix T_n (1,2)} {\bf T_E}^n(1,2) =
\left({\bf
T_E}^n\right)^{\dagger} (2,1) = 
\left[\begin{array}{cccccc}
c_0 & f_1 & 0 & \multicolumn{3}{c}{} \\
c_1 & c_0 & f_1 & 0 & \multicolumn{2}{c}{}  \\
c_2 & c_1 & c_0 & f_1 & 0 &   \\
0& c_2 & c_1 & c_0 & f_1 &  0  \\
\multicolumn{6}{c}{\dotfill} \\
\multicolumn{3}{c}{}  & c_2 & c_1 & c_0\\
\end{array}
\right]. \eeeq

\no Permutation matrix ${\bf P}^n$ is an orthonormal matrix
satisfying ${\bf P}^n \left({\bf P}^n\right)^T = {\bf I}_{2n}$
which enables us to rewrite the mutual information $
I^{(n)}_{E-MacA}$ as \be \label{eq: transformed mutual information
I_maca}
 I^{(n)}_{E-MacA} & = & \frac{1}{n} \log \mbox{det} \left [ {\bf I}_{2n} +
\frac{1}{{\mathcal N}_0}
\left[
\begin{array}{cc}
\Sigma_1 & {\bf 0}_n \\
{\bf 0}_n & \Sigma_2
\end{array}\right]
{\bf \mathcal T}^{(2n)}
\right] \nonumber \\
& = & \frac{1}{n} \log \mbox{det} \left [ {\bf I}_{2n} +
\mbox{SNR} \frac{2}{K+1}
{\bf \mathcal T}^{(2n)} \right] \nonumber \\
& = & \frac{1}{n} \sum_{k=1}^{2n} \log \left[ 1 + \frac{2\mbox{SNR}
}{K+1} \cdot \nu_k\left({\bf \mathcal T}^{(2n)} \right)
\right], \ee

\no where ${\bf 0}_n$ is a $n \times n$ zero matrix,  $\Sigma_j =
E \left[ \underline{\bf b}_{R_j} \underline{\bf b}^{\dagger}_{R_j}
\right] = \frac{2}{K+1} \hat{P}_s {\bf I}_n, \, j =1, 2$ and
$\mbox{SNR} = \frac{\hat{P}_s}{\mathcal{N}_0}$,   $\nu_k\left({\bf
\mathcal T}^{(2n)} \right)$ is the $k$th eigenvalue of the
$2\times 2$ block matrix ${\bf \mathcal T}^{(2n)}$. To obtain the
limit of $ I^{(n)}_{E-MacA}$ as $n$ goes to infinity,
 Theorem 3 in \cite{gazzah_01_block_toeplitz}
 regarding the eigenvalue distribution of Hermitian block
Toeplitz matrices can be  directly applied here  yielding 
the
following theorem:

\begin{theorem} \label{theoremtoeplitz}
As $n \rightarrow \infty$, we have
\beeq \label{eq: asymptotic distributions of block toeplitz matrix}
 \lim_{n \rightarrow \infty}  I^{(n)}_{E-MacA}  =
\frac{1}{2\pi} \int_{-
\pi}^{\pi} \sum_{j=1}^{2} \log \left[ 1 +\frac{2}{m} \mbox{SNR}
\cdot \nu_j \left( {\bf T_E}(\omega) \right) \right] \, d \omega,
\eeeq

\no where $ \nu_j \left( {\bf T_E}(\omega) \right)$ is the $j$th
largest eigenvalue of a Hermitian matrix
\beeq
\label{eq: matrix of frequency domain}
 {\bf T_E}(\omega)   =
   \left[ \begin{array}{cc}
|\alpha_{R_1,D}|^2 t_E^{(1,1)}(\omega) &  \alpha_{R_1,D} \alpha^*_{R_2,D} t_E^{(1,2)}(\omega) \\
\alpha^{*}_{R_1,D} \alpha_{R_2,D} t_E^{(2,1)}(\omega) &
|\alpha_{R_2,D}|^2 t_E^{(2,2)}(\omega)
\end{array}
\right], \eeeq

\no
 whose
entries $t_E^{(j,l)}(\omega)$ are the discrete-time Fourier
transforms of the elements of  Toeplitz matrices in ${\bf \mathcal
T}^{(2n)}$, i.e.
 $t_E^{(j,l)}(\omega) \stackrel{\bigtriangleup}{=}
\sum_{k} t_{E,k}\left(j,l \right) e^{-i k \omega }, \, j, l =1,
2$,  and are determined as
  \be
t_E^{(1,1)}(\omega) & = &\left[1 + a_1 e^{-i \omega} + a_1 e^{i \omega} \right],\nonumber \\
t_E^{(1,2)} (\omega) & = & \left[ c_1 e^{-i \omega} + c_2 e^{-i
2\omega} + c_0 + f_1 e^{i\omega} \right] =
\left(t_E^{(2,1)}(\omega)\right)^{*},
 \nonumber \\
t_E^{(2,2)}(\omega) & = &   \left[ 1 + d_1 e^{-i \omega} + d_1
e^{i \omega} \right]. \ee
\end{theorem}

\begin{proof}

 Theorem 3 in \cite{gazzah_01_block_toeplitz}
 regarding the eigenvalue distribution of Hermitian block
Toeplitz matrices yields the desired results.

\end{proof}

\begin{corollary}
For the relay channel model described in Section~\ref{section: system model},
suppose nodes $N_{R_1}$ and
$N_{R_2}$ employ the same waveform $s(t)$ such that $a_1 = d_1$ as defined in
(\ref{eq: coefficients a_1 and d_1}).
The limit of mutual information in Theorem~\ref{theoremtoeplitz} can thus
be further simplified as
\begin{gather} \label{eq: final approximation for I_{e-maca} for large n}
I_{E-MacA}  = 
\lim_{n\rightarrow \infty} I^{(n)}_{E-MacA}   \nonumber \\
 =   \frac{1}{2 \pi} \int_{-\pi}^{\pi} \log \left[ 1 + \rho_0
\frac{1}{2}\left( |\alpha_{R_1,D}|^2 + |\alpha_{R_2,D}|^2 \right)
\sum_{k=1}^2 \tilde{\nu}_k (\omega)
  \right.  \nonumber \\
  \left. + \rho^2_0
 |\alpha_{R_1,D}|^2  |\alpha_{R_2,D}|^2 \prod_{k=1}^2 \tilde{\nu}_k (\omega)
\right]. \end{gather}

\no where $ \sum_{k=1}^2 \tilde{\nu}_k (\omega) = 2
\left( 1 + 2 a_1 \cos \omega \right)$ and $ \prod_{k=1}^2
\tilde{\nu}_k (\omega) =\left[ \left( 1 + 2 a_1 \cos \omega
\right)^2 - |\hat{\rho}(\omega)|^2 \right]$, with
 $\hat{\rho}(\omega)= c_1 e^{-i \omega} + c_2 e^{-i
2\omega} + c_0 + f_1 e^{i\omega}$.

\end{corollary}

\begin{proof}

Eigenvalues of the $2\times 2$ matrix ${\bf T_E}(\omega)$ satisfy
the  following relationship \be \label{eq: relationship of the
ei-values of T(omega)} \sum_{j=1}^2 \nu_j \left( {\bf T_E}(\omega)
\right) & = & \frac{1}{2} \left( |\alpha_{R_1,D}|^2 +
|\alpha_{R_2,D}|^2 \right) \sum_{k=1}^2 \tilde{\nu}_k (\omega) \nonumber \\
\prod_{j=1}^2  \nu_j \left( {\bf T_E}(\omega) \right)  & = &
 |\alpha_{R_1,D}|^2  |\alpha_{R_2,D}|^2
\prod_{k=1}^2  \tilde{\nu}_k (\omega) , \ee

\no where $\tilde{\nu}_k (\omega), k=1,2$ are eigenvalues of a
Hermitian matrix \beeq \label{eq: Original h matrix} {\bf
\tilde{T}_E}(\omega)   =
   \left[ \begin{array}{cc}
 t_E^{(1,1)}(\omega) &   t_E^{(1,2)}(\omega) \\
 t_E^{(2,1)}(\omega) & t_E^{(2,2)}(\omega)
\end{array}
\right], \eeeq

\no and they satisfy  $ \sum_{k=1}^2 \tilde{\nu}_k (\omega) = 2
\left( 1 + 2 a_1 \cos \omega \right)$ and $ \prod_{k=1}^2
\tilde{\nu}_k (\omega) =\left[ \left( 1 + 2 a_1 \cos \omega
\right)^2 - |\hat{\rho}(\omega)|^2 \right]$, with
 $\hat{\rho}(\omega)= c_1 e^{-i \omega} + c_2 e^{-i
2\omega} + c_0 + f_1 e^{i\omega}$. Under these relationships and
Theorem~\ref{theoremtoeplitz}, we obtain (\ref{eq: final approximation for
I_{e-maca} for large n}).

\end{proof}

\subsubsection{Positive Definiteness of Matrix ${\bf \tilde{T}_E}(\omega)$ and
DM-tradeoff of Asynchronous Coded Scheme}

In this section,  we  show under certain conditions the Hermitian matrix
 ${\bf \tilde{T}_E}(\omega)$ defined in (\ref{eq: Original h matrix})
is positive definite for all $\omega \in [-\pi, \pi]$ and
consequently there exists a positive lower bound
$\lambda^{(2)}_{\mbox{min}}$ for eigenvalues $\tilde{\nu}_k
(\omega)$.
 As a result,  the DM-tradeoff of
this $2\times 1$ MISO system employing asynchronous space-time
codes is equal to  that of a parallel frequency flat fading
channel with two independent users.

\begin{theorem}
\label{theorem: positive defininity} When a time-limited
waveform $s(t)=0, t \notin [0, 2 T_s]$ is chosen such that complex
signals $F_1(t,\omega) =  \sum_{k=0}^2 s(t+kT_s) e^{j k\omega}$
and $F_2(t,\omega) = \sum_{k=0}^2 s(t-\tau+kT_s) e^{j k\omega}$
are linearly independent with respect to $t \in [0, T_s]$ for any
  $\omega \in [-\pi, \pi]$, the matrix
 ${\bf \tilde{T}_E}(\omega)$ is always positive definite for $\forall \omega
\in [-\pi, \pi]$ and there exists  positive numbers
$\lambda^{(2)}_{\mbox{min}}>0$ and $0 < \lambda^{(2)}_{\mbox{max}}
\leq 10$ such that $\lambda_{\mbox{min}}(\omega) \geq
\lambda^{(2)}_{\mbox{min}}$ and $\lambda_{\mbox{max}}(\omega) \leq
 \lambda^{(2)}_{\mbox{max}}$,
where $\lambda_{\mbox{min}}(\omega) $ and
$\lambda_{\mbox{max}}(\omega)$ are the   minimum  and maximum
eigenvalues of the matrix ${\bf \tilde{T}_E}(\omega)$,
respectively.
\end{theorem}

\begin{proof} See  Appendix~\ref{positiveproof}.
 As shown in the Appendix~\ref{positiveproof}, a similar
conclusion  can be reached when $s(t)$ spans over an arbitrary number
of finite symbol periods, i.e. $s(t) =0, t \notin [0, M T_s]$,
$M\geq 1$.

\end{proof}

If  $s(t)$ satisfies  the condition in Theorem~\ref{theorem:
positive defininity}, we can upper- and lower-bound the mutual
information  $I_{E-MacA} $ in (\ref{eq: final approximation for
I_{e-maca} for large n}) through bounding eigenvalues
$\tilde{\nu}_k (\omega), k=1,2$ of
 ${\bf \tilde{T}_E}(\omega)$.
The lower bound of $I_{E-MacA}$ is
\begin{gather} \label{eq: lowerbound of I-Emaca}
I_{E-MacA} \geq   \frac{1}{2 \pi} \int_{-\pi}^{\pi} \log \left[ 1 + \rho_0
\left( |\alpha_{R_1,D}|^2 + |\alpha_{R_2,D}|^2 \right)
 \lambda^{(2)}_{\mbox{min}} \right.  \nonumber  \\
\left. + \rho^2_0
 |\alpha_{R_1,D}|^2  |\alpha_{R_2,D}|^2 \left( \lambda^{(2)}_{\mbox{min}} \right)^2
\right] \nonumber \\
 =   \sum_{k=1}^2 \log \left[ 1 + \rho_0
|\alpha_{R_k,D}|^2  \lambda^{(2)}_{\mbox{min}} \right]
\stackrel{\bigtriangleup}{=} I^{(L)}_{E-MacA}.
 \end{gather}

\no Similarly, we can upper bound $I_{E-MacA}$ by \beeq
\label{eq: upperbound of I-Emaca} I_{E-MacA} \leq \sum_{k=1}^2
\log \left[ 1 + \rho_0 |\alpha_{R_k,D}|^2
\lambda^{(2)}_{\mbox{max}} \right] \stackrel{\bigtriangleup}{=}
I^{(U)}_{E-MacA}. \eeeq

\no The upper-bound is not surprising since it means the
performance of a $2\times 1$ MISO system is bounded from above by
that of a MIMO system with two completely separated channels.

The
fundamental reason behind the lower bound is because the matrix
${\bf \tilde{T}_E}(\omega)$ is positive definite for arbitrary
$\omega \in [-\pi, \pi]$. This enables the channel of large block length
as characterized by (\ref{eq: discrete time extended asynchro.
model})  has mutual information at least as
large as that of a two-user parallel Rayleigh fading channel, which takes a
form of  $\sum_{k=1}^2 \log\left(1 +  \rho_0 \kappa |\alpha_{R_k,D}|^2 \right)
$, where $\kappa$ is a positive constant. Different finite values taken by
$\kappa$, e.g. either $\lambda^{(2)}_{\mbox{max}}$ or
$\lambda^{(2)}_{\mbox{min}}$, have no effect on the
 diversity-multiplexing  tradeoff function.
Therefore, the channel between two relay nodes and the destination
when asynchronous space-time coding  is employed is
equivalent to a two-user parallel  fading channel in terms of the
diversity-multiplexing tradeoff. This result is summarized by
Lemma~\ref{theorem_parallel}.

\begin{lemma} \label{theorem_parallel}
When both relay nodes succeed in decoding the source information and employ asynchronous
space-time codes across them, the outage probability
$\mbox{Pr} \left[ I_{E-MacA} < R,\, |\mathcal{D}(s)|=2 \right]$
behaves asymptotically as
\begin{gather}
\mbox{Pr} \left[ I_{E-MacA} < R,\, |\mathcal{D}(s)|=2 \right] \nonumber \\ \sim
\mbox{Pr} \left[ \sum_{k=1}^2 \log\left(1 +  \rho_0 \kappa |\alpha_{R_k,D}|^2 \right) < R\right],
\end{gather}
\no where $\kappa$ is a positive constant.

\end{lemma}

\begin{proof}

The proof is straightforward using lower bound and upper bound of $I_{E-MacA}$
in (\ref{eq: lowerbound of I-Emaca} ) and  (\ref{eq: upperbound of I-Emaca}),
respectively.

\end{proof}

The overall outage probability counting the direct link between source and its
destination, as well as the relay-destination link when $|\mathcal{D}(s)|=2$,
can also be determined in a similar manner.

\begin{theorem} \label{theoremAsyn}
Given asynchronous space-time codes are deployed by relay
 nodes when $|\mathcal{D}(s)|=2$,  the conditional outage probability of
$\mbox{Pr} \left[ I_{A-stc} <R | |\mathcal{D}(s)|=2  \right]$ has
an asymptotic equivalence the same as that of a parallel channel
with $3$ independent paths, i.e. 
\begin{gather} \mbox{Pr}\left[ I_{A-stc} <
R | |{\mathcal D}(s)| = 2 \right] \nonumber \\ \sim
\widetilde{\mbox{SNR}}^{-\left( 3 - 2 r \right)} \cdot 2  \left(r
\log \widetilde{\mbox{SNR}} \right)^2 \prod_{ k\in \left\{S, R_1,
R_2 \right\}} {\tilde\lambda}_{k,D}, \end{gather}

\no if a time limited waveform $s(t) = 0, t \notin [0, 2 T_s]$
satisfying the condition outlined in Theorem~\ref{theoremtoeplitz}
is employed.

\end{theorem}

\begin{proof}

To study the overall DM-tradeoff given $|\mathcal{D}(s)|=2$, we also need to
bound $I_{E-SD}$ in
 (\ref{eq: the first whole link mutual infor. for
asyn.s-time codes}).  By making $\alpha_{R_2,D}=0$ in
(\ref{eq: final approximation for I_{e-maca} for large n}), we
obtain
 \begin{gather} \label{eq: direct channel mutual infor.}
I_{E-SD}  =   \frac{1}{2 \pi} \int_{-\pi}^{\pi} \log \left[ 1 +
\rho_0   |\alpha_{S,D}|^2  \left( 1 + 2 a_1 \cos \omega \right)
\right] \, d \omega  \nonumber \\
 =   \log \left[ 1 +\rho_0  |\alpha_{S,D}|^2 \right] + \frac{1}{2\pi}
\int_{-\pi}^{\pi}
\log\left[ 1 + \frac{2 \rho_0 |\alpha_{S,D}|^2  a_1 }{1+\rho_0
|\alpha_{S,D}|^2} \cos \omega \right] \, d \omega \nonumber  \\
 =  \log \left( 1+ \rho_0 |\alpha_{S,D}|^2 \right) +
\log\left[{1 + \sqrt{1 - \left( \frac{2 \rho_0  |\alpha_{S,D}|^2 a_1
}{1 +\rho_0  |\alpha_{S,D}|^2 }\right)^2}}\right] -1
\end{gather}

\no where the last equality is based on  the integral equation
(\ref{eq: table integral}). Since $\sum_{k=1}^2 \tilde{\nu}_k
(\omega) = 2\left( 1+ 2 a_1 \cos\omega \right)>0$, it always holds for $a$  to 
satisfy $|a|<1/2$, which  justifies  the second  equation above.
Therefore, the bounds of $I_{E-SD}$ are 
\be \label{I-E-SD bounds}
 I_{E-SD}^{(L)}  & \stackrel{\bigtriangleup}{=}& \log \left( 1+ \rho_0
|\alpha_{S,D}|^2 \right) -1 <  I_{E-SD} \nonumber \\
& \leq &   \log \left( 1+ \rho_0 |\alpha_{S,D}|^2 \right)
  \stackrel{\bigtriangleup}{=} I_{E-SD}^{(U)}.
\ee

Bounds on $I_{E-MacA}$ and $I_{E-SD}$ as shown in (\ref{eq:
lowerbound of I-Emaca}), (\ref{eq: upperbound of I-Emaca}) and
(\ref{I-E-SD bounds}), respectively, can thus yield  bounds on the
whole link outage probability  $\mbox{Pr} \left[ \frac{1}{2}
\left( I_{E-SD} + I_{E-MacA}\right) < r \log \mbox{SNR} \right]$ when relays
are all in $\mathcal{D}(s)$. 
Comparing these bounds, we can conclude the lower and upper bounds
of the overall outage probability has the same order of
diversity-multiplexing tradeoff as a system with $3$ parallel independent
Rayleigh fading channels whose mutual information takes the form of
$\frac{1}{2} \sum_{j= \in \left\{S,R_1, R_2 \right\}} \log \left[
1 + \rho_0 |\alpha_{j,D}|^2 \right]$. Hence, when the decoding set
includes both relay nodes, the overall outage probability has the
following asymptotic equivalence,

\begin{gather} \label{eq: asymptotic equivalence of  I-EMACA}
\mbox{Pr}\left[ I_{A-stc} < R | |{\mathcal
D}(s)| = 2 \right] \nonumber \\ \sim
\mbox{Pr} \left[  \frac{1}{2} \sum_{j= \in \left\{S,R_1,
R_2 \right\}}  \log \left[ 1 + \rho_0 |\alpha_{j,D}|^2 \right]
< R(\mbox{SNR}) \right].
\end{gather}

Following the same approach as in Section \ref{section: syn.
distc},  we obtain
\smallskip
\begin{gather}  \label{eq: asympt. result on I-EMACA}
\mbox{Pr} \left[  \sum_{j= \in \left\{S,R_1,
R_2 \right\}}  \frac{1}{2} \log \left[ 1 + \rho_0 |\alpha_{j,D}|^2 \right]
< R \right]
 \sim  \nonumber \\
\mbox{Pr} \left[ \widetilde{\mbox{SNR}}^{\sum_{i \in \left\{S,R_1, R_2 \right\}} (1- \beta_{i,D})^{+}} < \widetilde{\mbox{SNR}}^{2r} \right] \nonumber \\
={ \int_{ \beta_{k,D}  \in {\hat A} }  \left( \log
\widetilde{\mbox{SNR}} \right)^3 \prod_{k \in \{S, R_1, R_2\}}
\widetilde{\mbox{SNR}}^{- \beta_{k,D}}
\tilde{\lambda}_{k,D}}  \nonumber \\
  \exp\left\{ -\tilde{\lambda}_k
\widetilde{\mbox{SNR}}^{- \beta_{k,D}} \right\} \, d \beta_{k,D} \nonumber \\
  \hspace{-0.5in} \sim \int_{ \beta_{k,D}  \in {\hat A} }
\left( \log \widetilde{\mbox{SNR}} \right)^3 \prod_{k \in \{S,
R_1, R_2\}} \widetilde{\mbox{SNR}}^{- \beta_{k,D}}
\tilde{\lambda}_{k,D} \, d \beta_{k,D}  \nonumber \\
  \hspace{-1.0in} \sim \widetilde{\mbox{SNR}}^{-\left( 3 - 2 r
\right)} \cdot 2  \left(r \log \widetilde{\mbox{SNR}} \right)^2
\prod_{ k\in \left\{S, R_1, R_2 \right\}} {\tilde\lambda}_{k,D},
\end{gather}

\no where \beeq {\hat A} = \left\{ \sum_{k\in \left\{ S, R_1, R_2
\right\}} \left( 1 -  \beta_{k,D} \right)^+ < 2r,\,  \beta_{k,D}
\geq 0  \right\}, \eeeq

\no and  the last asymptotic relationship is obtained similarly as
 (\ref{eq: outage prob. asymp. with one relay node}). Combining (\ref{eq: asymptotic equivalence of  I-EMACA})
 and (\ref {eq: asympt. result on I-EMACA}) thus completes the proof of Theorem~\ref{theoremAsyn}.

 \end{proof}

Therefore, if $s(t)$ lasting for two symbol periods satisfies the
condition in Theorem \ref{theorem: positive defininity},
and two relay nodes both successfully decode the source codewords,
the rate of the outage probability approaching zero as
$\mbox{SNR}$ goes to infinity is $\mbox{SNR}^{-3+2r}, \, r \in [0,
1/2]$ which is better than $\mbox{SNR}^{-3+4r}$ in Lemma~\ref{lemma3}.
This result explicitly demonstrates the benefit of employing
asynchronous space-time codes under the  presence
of relay asynchronism in terms of DM-tradeoff.

Having obtained the asymptotic behavior of outage probability when
two relay nodes are both in the decoding set $\mathcal{D}(s)$, we
now shift our focus towards  the overall
 DM-tradeoff averaged over all possible
outcomes of $\mathcal{D}(s)$.  We prove next that the overall
DM-tradeoff is $d_{A-stc}(r)=3-6r$ which is equal to  that for
both independent coding based distributed delay diversity and
synchronous space-time coded cooperative diversity schemes.

\begin{theorem}
\label{theoremAsyn_overall}  When the  time-limited waveform
$s(t)=0, t \notin [0, 2 T_s]$ satisfies  conditions specified in
Theorem~\ref{theorem: positive defininity},
 the DM-tradeoff of asynchronous space-time-time coded approach is
 \be \label{eq: DM tradeoff of A-STC}
 d_{A-stc}(r)&  = &
\lim_{ \mbox{SNR} \rightarrow \infty} - \frac{ \log
\left( \mbox{Pr} \left[ I_{A-stc} < R \left( \mbox{SNR} \right)
\right] \right)}{\log \mbox{SNR}} \nonumber \\
&  = &  3 ( 1 - 2 r) = d_{stc}(r), \, 0 \leq r < 1/2.
\ee
\end{theorem}

\begin{proof}

When no relay succeeds in decoding or only one of two relay nodes
has decoded correctly, the overall capacity takes the form of
either $I_{A-stc} = I_{E-SD}/2$ or $I_{A-stc} = \left[I_{E-SD} +
I_{E-RD}\right]/2$, where $I_{E-SD}$ was obtained in (\ref{eq:
direct channel mutual infor.}) and $I_{E-RD}$ has a similar
expression as $I_{E-SD}$ except fading variable $\alpha_{S,D}$ is
substituted by $\alpha_{R,D}$ in (\ref{eq: direct channel mutual
infor.}).

 We can therefore infer based on lower and upper bounds in
(\ref{I-E-SD bounds}) that the conditional outage probability
$\mbox{Pr}\left[I_{A-stc}< r \log \mbox{SNR}, |\mathcal{D}(s)| =j
\right]$  has the asymptotic term determined by
$\mbox{SNR}^{-(3-6r)}$ and $\mbox{SNR}^{-(3-4r)}$ for $j=0$ and
$j=1$, respectively, which are the same as  both  synchronous
space-time coded  and independent coding based distributed delay
diversity schemes.

 Meanwhile, the vanishing rate of
$\mbox{Pr}\left[I_{A-stc}< r \log \mbox{SNR}, |\mathcal{D}(s)| =2
\right]$ towards zero is subject to $\mbox{SNR}^{-(3-2r)}$,  as
demonstrated by  Theorem~\ref{theoremAsyn}. However, the
performance improvements using asynchronous space-time codes
across two relays is not going to be reflected in the overall
DM-tradeoff function because the dominant term among
$\mbox{SNR}^{-(3-6r)}$, $\mbox{SNR}^{-(3-4r)}$ and
$\mbox{SNR}^{-(3-2r)}$ for $ r \in [0, 1/2]$ is
$\mbox{SNR}^{-(3-6r)}$. Consequently, we conclude the  overall
DM-tradeoff is $d_{A-stc}(r)=3-6r$ and thus complete the proof of
Theorem~\ref{theoremAsyn_overall}.

\end{proof}

\subsubsection{Comparison with Synchronous Approach Under
Arbitrary $\mbox{SNR}$}

In order to further demonstrate the benefits of completely
exploiting spatial and temporal degrees of freedom by using
asynchronous space-time codes, we investigate the performance
improvements in terms of achievable rate for the channel between
two relay nodes and destination under an arbitrary finite
$\mbox{SNR}$. We restrict our attentions to a particular case when
the baseband waveform $s(t)$ is limited within one symbol period,
i.e. $s(t) = 0$ for $t \notin [0,T_s]$.

\begin{theorem}\label{theoremFiniteSNR}

If $s(t)$ is time-limited within one symbol period and selected to
make ${\bf \tilde{T}_E}(\omega)$ a positive definite matrix for
all $\omega \in [-\pi, \pi]$ in (\ref{eq: final approximation for
I_{e-maca} for large n}), the mutual information rate between two
relay nodes and destination is strictly greater than that with
synchronous space-time coded approach for any $\mbox{SNR}$, i.e.
\beeq I_{E-MacA} >  \log \left[ 1 + \rho_0 \left(
|\alpha_{R_1,D}|^2 + |\alpha_{R_2,D}|^2 \right) \right] = I_{STC},
 \eeeq
\no for any $\mbox{SNR}$.

\end{theorem}

\begin{proof}

 Consider the term $\sum_{k=1}^2 \tilde{\nu}_k (\omega)$
 in (\ref{eq:
final approximation for I_{e-maca} for large n}) which is the sum
of eigenvalues of the matrix ${\bf \tilde{T}_E}(\omega)$
satisfying
  $\sum_{k=1}^2 \tilde{\nu}_k (\omega) =
\mbox{Trace} \left({\bf \tilde{T}_E}(\omega)\right)$. If $s(t)$ is
time-limited within one symbol period and selected to make ${\bf
\tilde{T}_E}(\omega)$ a positive definite matrix for all $\omega
\in [-\pi, \pi]$, we have  $\mbox{Trace} \left({\bf
\tilde{T}_E}(\omega)\right) =2$ and $\tilde{\nu}_k (\omega)>0$, as
shown in Appendix~\ref{positiveproof}.
 Under these conditions, we
obtain 
\begin{gather} \label{eq: lower bound of I-EMACA} I_{E-MacA}    \nonumber
\\
> \frac{1}{2 \pi} \int_{-\pi}^{\pi} \log \left[ 1 + \frac{\rho_0}{2}
\left( |\alpha_{R_1,D}|^2 + |\alpha_{R_2,D}|^2 \right)
\sum_{k=1}^2 \tilde{\nu}_k (\omega)
\right] \nonumber \\
 =  \log \left[ 1 + \rho_0 \left( |\alpha_{R_1,D}|^2 +
|\alpha_{R_2,D}|^2 \right) \right] = I_{STC},  \end{gather}

\no which demonstrates  $I_{E-MacA}$ is  strictly larger than the
capacity of a $2 \times 1$ MISO system employing synchronous
space-time codes in a frequency flat fading channel, i.e. asynchronous
space-time codes increases the capacity of the MISO system.

\end{proof}

If $s(t)$ is a truncated squared-root-raise-cosine waveform
spanning over  $M >1$ symbol periods with  $M \in \mathcal{Z}^+$,
it has been shown in Appendix~\ref{positiveproof} that if
$\mbox{Trace} \left({\bf \tilde{T}_E}(\omega) \right) \approx 2 $
and  $\tilde{\nu}_k (\omega)>0$ for   some $M$ and $s(t)$,
 a similar result as (\ref{eq: lower bound of I-EMACA})
can be obtained as well, 
\begin{gather} \label{eq: lower bound of I-EMACA for SRRC} 
I_{E-MacA} 
> \nonumber \\
 \frac{1}{2 \pi} \int_{-\pi}^{\pi} \log \left[ 1 +
\frac{\rho_0}{2} \left( |\alpha_{R_1,D}|^2 + |\alpha_{R_2,D}|^2
\right) \sum_{k=1}^2 \tilde{\nu}_k (\omega)
\right] \nonumber \\
 \approx  \log \left[ 1 + \rho_0 \left( |\alpha_{R_1,D}|^2 +
|\alpha_{R_2,D}|^2 \right) \right]. \end{gather}

Of course,  when $M$ increases, the memory length of the
equivalent vector ISI channel increases as well,  as shown by Eq.
(\ref{eq: discrete time extended asynchro. model}), which
naturally  increases the decoding complexity.
 This  manifests the cost incurred for
having a  better  diversity-multiplexing tradeoff and higher
mutual information than the synchronous space-time-coded scheme.
Therefore, a time-limited root-mean-squared  (RMS) waveform
lasting for only  one symbol period  is preferred under the
bandwidth constraint \cite{roger_91}.

\subsubsection{Extensions to N-Relay Network}

Although the  channel model we have focused on in this paper 
concerns only with   two relay nodes, the methodologies and major ideas
behind our approaches to attaining DM-tradeoff can be applied to
cases of relay network with $N>2$ relay nodes.

For example, when asynchronous space-time code is employed across
$N\geq M>2$ relay nodes, the mutual information between these $M$
active relay nodes and destination can be obtained using the
similar technique in proving Theorem~\ref{theoremtoeplitz}.  In
addition, similar conditions as in Theorem~\ref{theorem: positive
defininity} under which we have strictly positive definite matrix
can be developed
 as in \cite{roger_92} such that we can also bound the mutual
 information as we did in  (\ref{eq: lowerbound of I-Emaca}) and
(\ref{eq: upperbound of I-Emaca}). Consequently, we can foresee
the relay-destination link is equivalent to a parallel channel
with $M$ independent links in terms of DM-tradeoff function. As
for the overall DM-tradeoff function after averaging out all
possible outcomes of decoding  set of relay nodes, we will arrive
at the same conclusion as two-relay network due to the same
bottleneck caused by an empty decoding set.

\subsection{ Bottleneck Alleviation with Mixing Approach}
\label{sec: bottle neck}

As demonstrated in Section~\ref{section: Delay Diversity}
and Section~\ref{section: asyn. coding}, there exists a bottleneck case
dominating the overall DM-tradeoff function. This is mainly caused by the slowly
vanishing rate of the
outage probability when no relay node succeeds in decoding the
source packets, and consequently the destination node only has access to the
packets sent by source directly.
For all schemes we have proposed,  we 
assume an orthogonal channel
allocation strategy in which source transmits only in the first phase
and
relays forward packets after they decode the source messages
correctly in the second phase. This orthogonal channel allocation
is the fundamental reason of why the valid range of multiplexing gain $r$
is confined over an interval $[0,1/2]$. 

To address the aforementioned  issue of restricted  multiplexing gain,
 Dynamic Decode and Forward (DDF) and Non-orthogonal
Amplify and Forward (NAF) schemes are proposed in \cite{gamal_submitted},
through which the overall
DM-tradeoff is improved . Both of these two
schemes allow source to continuously  transmit during an entire
frame. In the DDF scheme, relays do not forward until they collect
sufficient energy to decode the source signals. In the NAF scheme,
relays forward the scaled received source signals in  alternative
intervals. The resulting overall DM-tradeoff of these schemes are
 \beeq \label{eq: NAF} d_{NAF}(r) = (1-r) +
K \left( 1- 2r \right)^+, \, 0 \leq r \leq 1, \eeeq

\no and \beeq \label{eq: DDF} d_{DDF} (r) =\left\{
\begin{array}{ll}
(K+1) (1-r) & 0 \leq r \leq \frac{1}{K+1} \\
1 + \frac{K (1-2r)}{1-r} & \frac{1}{K+1} \leq  r \leq \frac{1}{2}\\
\frac{1-r}{r} & \frac{1}{2} \leq r \leq 1
\end{array},
\right. \eeeq

\no where $K$ is the number of relay nodes in the system and
$x^+ =\max(x,0)$.

\begin{figure}[h]
\begin{center}
\includegraphics[scale=0.45]{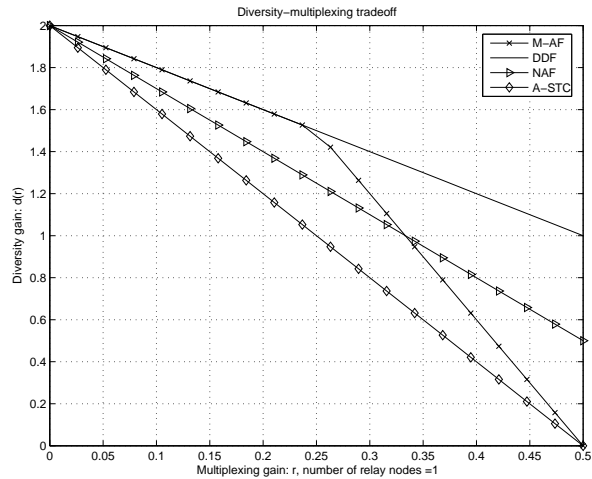}
\caption{Diversity-multiplexing tradeoff of cooperative diversity schemes. There is one relay node
between the source and its destination. Diversity gains $d_{M-AF,N=2}(r)$, $d_{DDF}(r)$ and $d_{NAF}(r)$
are obtained based on
(\ref{eq: diversity AF N=2}), (\ref{eq: DDF}) and
(\ref{eq: NAF}) for $N=2$, respectively, and $d_{A-stc}(r)=2-4r$.
}
\label{fig_2}
\end{center}
\end{figure}

\begin{figure}[h]
\begin{center}
\includegraphics[scale=0.45]{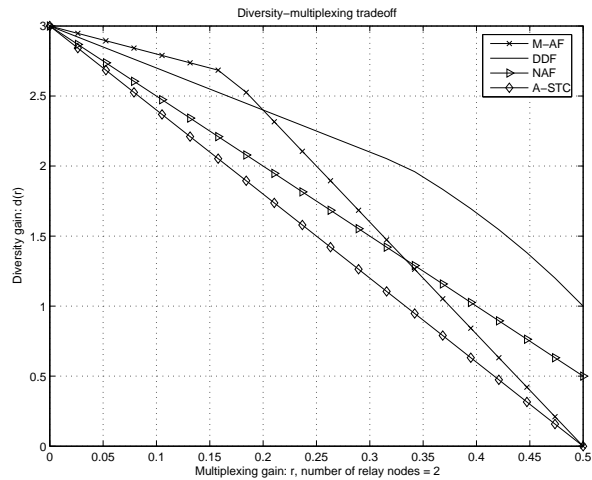}
\caption{Diversity-multiplexing tradeoff of cooperative diversity schemes. There are two relay nodes
between the source and its destination. Diversity gains $d_{M-AF,N=3}(r)$, $d_{DDF}(r)$ and $d_{NAF}(r)$
are obtained based on
(\ref{eq: 2-relay node M-AF}),  (\ref{eq: DDF}) and
(\ref{eq: NAF}) for $N=3$, respectively, and $d_{A-stc}(r)=3-6r$ is obtained in
Section~\ref{section: asyn. coding}.
}
\label{fig_3}
\end{center}
\end{figure}

\subsubsection{One-Relay Case}

First suppose there is only one relay node between  $N_S$ and $N_D$ and
there are two phases in transmission as assumed in Section~\ref{section:
system model}.  The proposed mixing strategy  works as follows. Assume
the channel fading parameter $\alpha_{S,R}$ can be measured perfectly at
a relay node such that it can determine whether there will be an outage
given  current channel realizations. If there is no outage, the relay
node works similarly as described in previous sections by performing
decode-and-forward; otherwise, instead of dropping the received source
packets, relay amplify-and-forwards the incoming source signals with
an amplifying  coefficient $\beta= \sqrt{\frac{P}{P |\alpha_{S,R}|^2 +
N_0}}$ to maintain its constant transmission power. It turns out the
overall diversity-multiplexing tradeoff can be improved by this simple
mixing scheme as shown next.

It has been proved  in \cite{laneman_02}  that the AF and
selection decode-and-forward schemes for a single relay network
have the same DM-tradeoff function: $d_{AF}(r)=d_{DF} = 2 (1-2r)$,
for $ r \in [0,1/2]$. Applying the similar analytical approach as
in Section \ref{section: syn. distc}, the outage probability for a
relay channel with only one relay node performing the
decode-and-forward has an asymptotic equivalence consisting of two terms:
  \beeq
\label{eq: single relay
diversity} P_{out} \sim A \cdot \mbox{SNR}^{-(2-2r)} + B \cdot
\mbox{SNR}^{-2(1-2r)}, \eeeq

\no where the first term is contributed by relay's successful
decoding and then independent encoding over successive two phases, the
second term is due to relay's  dropping
of the received signals because of its failure in decoding phase,
$A$ and
$B$ are some finite constants. Therefore, the overall DM-tradeoff
is $d_{DF}(r)=2-4r$ due to  the dominance of the slope $2-4r$
for $r \in [0,1/2]$.

Under the proposed mixing strategy, the slope
in the first term of (\ref{eq: single relay diversity}) is not
affected when  relay succeeds in decoding. The second term is,
however, changed to $\mbox{SNR}^{-(1-2r)-(2-4r)}$, where $(1-2r)$
is the slope characterizing the vanishing rate of the probability of
$|\mathcal{D}(s)|=1$ as derived in (\ref{eq: asymptotic of decoding set}) in
Section \ref{section: syn. distc}, and $(2-4r)$ is the slope for
the AF scheme.  Therefore, the mixing scheme has an overall
DM-tradeoff  \beeq \label{eq: diversity AF N=2}d_{M-AF, K=1}(r) =
\left\{\begin{array}{ll}
2-2r,   &  0 \leq r \leq 1/4 \\
3-6r, & 1/4< r \leq 1/2
\end{array}
\right.
\eeeq

\no  which is strictly greater than $d_{DF}(r) = 2-4r$ for any  $r \in (0,
1/2)$, and thus shows the advantage  of mixing  the
 amplify-and-forward scheme with the decode-and-forward scheme.

When $K=1$, the DM-tradeoff of NAF is $d_{NAF}(r) = 2 -3 r, \, 0 \leq r \leq
1/2$ from (\ref{eq: NAF}). It
shows NAF is dominated by  M-AF for $ 0
\leq r \leq 1/3$. As for the DDF scheme, the diversity gain is
$d_{DDF}(r) = 2(1-r) \geq d_{M-AF}(r)$. The preceding comparison is
illustrated by Figure~\ref{fig_2}.

\subsubsection{Two-Relay Case}

In this section,  we generalize the idea of mixing strategy   to
a two-relay case, where we show mixing approach  can even
outperforms the DDF scheme for some subset of  multiplexing gain $r$. The
result is stated in the following Theorem:

\begin{theorem}\label{theorem: two-relay Mix}
The overall DM-tradeoff
 $d_{M-AF, K=2}(r)$ of a two-relay channel under our proposed
mixing strategy  is

 \beeq \label{eq: 2-relay node M-AF} d_{M-AF, K=2}(r) = \left\{
\begin{array}{ll}
3 -2r, & 0  \leq r \leq  \frac{1}{6}\\
4-8 r, & \frac{1}{6} \leq r \leq \frac{1}{2}.
\end{array}
\right.
\eeeq

\end{theorem}

\begin{proof}

The proof relies on the mixing protocol which exploits the DM-tradeoff for
asynchronous cooperative diversity schemes studied in Section~\ref{section:
Delay Diversity} and Section~\ref{section: asyn. coding}.
The mechanism of the proposed  protocol for this 2-relay node M-AF scheme is
subject to the outcome of decoding at two relay nodes.

When both relay nodes fail in decoding i.e. $|\mathcal{D}(s)|=0$, only one of
them employs
AF and another one drops the received signals. In this case,
the conditional outage probability has $P_{out|
|\mathcal{D}(s)|=0} \sim \mbox{SNR}^{-2(1-2r)-(2-4r)}$, where
$2(1-2r)$ is the absolute slope of the probability of $\{ |\mathcal{D}(s)|
=0 \}$ and $(2-4r)$ is the slope of the outage probability under AF.

If $|\mathcal{D}(s)|=1$, WLOG, suppose  $N_{R_2}$ fails
and $N_{R_1}$ succeeds in decoding. Thereafter, $N_{R_1}$
performs decode-and-forward employing a
 complex Gaussian codebook independent of the source codebook,
  while   node 2 applies AF forwarding  a scaled copy of the received signal.
The outage probability given one node is in the decoding set
has an asymptotic equivalence $P_{out| |\mathcal{D}(s)|=1} (M-AF, K=2) \sim
\mbox{SNR}^{-(1-2r)-l_0(r)}$, where $(1-2r)$ is the slope for the
probability of  $\{ |\mathcal{D}(s)| =1\}$ and $l_0(r) $
represents the vanishing  rate of the outage probability in an
equivalent channel between $N_S$ and $N_D$ across two relay  nodes.
Next, we look into the bounds on
$l_0(r)$ under different assumptions on  the relative delay $\tau$ and show
$3-6r \leq l_0(r) \leq 3-4 r$.

If the relative delay $\tau$ between two relays is in the
order of an integer number of symbol periods,  since $N_{R_2}$
employs the same codewords as the source which is independent of
what $N_{R_1}$ transmits, the slope $l_0(r)$ is expected to lie  between
that of the repetition  coding based distributed delay diversity and
independent coding based delay diversity schemes, which are $3-6r$ and
$3-4r$, respectively,  as derived in Section \ref{section: Delay
Diversity}. Therefore,we have  $3- 6r \leq l_0(r)
\leq 3-4r $  in this case.

If $|\tau|/T_s$ is a non-integer and  $s(t)$ satisfies the
 condition specified in Theorem \ref{theorem: positive defininity},
the relay-destination link is equivalent to a two-user
parallel flat fading channel in terms of DM-tradeoff. Consequently,
the mutual information
of the entire link in this case has an asymptotic equivalence the same  as
$\frac{1}{2}\left[I_{AF}  +  \log \left( 1+ \rho_0
|\alpha_{R_2,D}|^2\right)\right]$,
 where  $I_{AF}$ is the mutual
information for an AF scheme taking the form of  $\log\left[1+ \rho_0
(|\alpha_1|^2 + |\alpha_2|^2) \right]$ as shown in \cite{laneman_02},
 where $\alpha_1$ and $\alpha_2$ are independent complex Gaussian random
variables.
Therefore, we obtain $l_0(r) = 3-4r$, the asymptotic term  characterizing the
vanishing rate of  the synchronous space-time-coded diversity scheme when two
relay nodes are both in the decoding set,  as determined by
Lemma~\ref{lemma3}.

From the  preceding analysis we obtain 
 $ 3-6r \leq l_0(r) \leq 3-4r$, which  leads us to  
\be
\mbox{SNR}^{-(4-8r)} & \stackrel{<}{\sim} & 
  P_{out| |\mathcal{D}(s)|=1}(M-AF, N=3) \nonumber \\ 
& \stackrel{<}{\sim} &
\mbox{SNR}^{-(4-6r)}, \, r \in [0, 1/2]. \ee

 If two relay nodes both succeed in decoding i.e.
$|\mathcal{D}(s)|=2$, the overall DM-tradeoff is equal to the
asynchronous space-time-coded cooperative diversity approach yielding
  $P_{out| |\mathcal{D}(s)|=2} \sim \mbox{SNR}^{-(3-2r)}$
under $\tau \in (0, T_s)$ and $s(t)$ satisfying
the condition in Theorem \ref{theorem: positive defininity}.

Putting all cases  together, we can determine
 the overall DM-tradeoff averaged over all possible outcomes of
the decoding set $\mathcal{D}(s)$, which is  subject to the dominant term
among $\{ \mbox{SNR}^{-(4-8r)}, \mbox{SNR}^{-(4-6r)},$ $
\mbox{SNR}^{-(3-2r)}\}$ subject to $r$. For $r \in [0, 1/6]$, $\mbox{SNR}^{-(3-2r)}$
is the slowest one, hence,
$d_{M-AF, N=3}(r) = 3-2r$; for $r \in (1/6, 1/2]$, $\mbox{SNR}^{-(4-8r)}$
is the dominant one, we have
 $d_{M-AF,
N=3}(r) = 4-8r $. We thus complete the proof of 
Theorem~\ref{theorem: two-relay Mix}.

\end{proof}

From  this case study,  we can conclude the  mixing strategy does improve
the DM-tradeoff over the  pure decode-and-forward approach having
$d_{A-stc}=3-6r$.  Moreover, comparing (\ref{eq: 2-relay node M-AF})
with (\ref{eq: DDF}) and (\ref{eq: NAF}) for $K=2$, we find the proposed
mixing strategy outperforms DDF and NAF for $r \in [0, 1/5]$, and $r \in
[0,1/3]$, respectively,  as shown in Figure~\ref{fig_3}. This observation
demonstrates  in order to improve the overall DM-tradeoff for cooperative
diversity schemes in relay channels, we need to consider approaches  which
not only relax the restriction on  sources transmitting   only half of
the total degrees of freedom as DDF and NAF in \cite{gamal_submitted},
but also exploit  advantages of employing asynchronous coded schemes as
demonstrated above using the proposed mixing strategy.


\begin{table*}
\centering
\begin{tabular}{|l|l|l|l|l|l|} \hline
$|\mathcal{D}(s)|$ & S-STC & ICB-DD & RCB-DD & ICB-DD-L & A-STC \\ \hline
 $0$, $ 0 \leq r \leq 1/2$ & $3-6r$ & $3-6r$ & $3-6r$ & $3-6r$ & $3-6r$  \\
\hline
 $1$ & $3-4r$ & $3-4r$ & $3-4r$ & $3-4r$ & $3-4r$  \\ \hline
 $2$ & $3-4r$ & $\in [3-6r, 3-4r]$ & $ \in [3-6r/\Delta_1, 3-6r]$&  $3-4r$ &
$3-2r$ \\ \hline \hline
Overall DM-tradeoff & $3-6r$ & $3-6r$ &  $\in [3-6r/\Delta_1, 3-6r]$ & $3-6r$
& $3-6r$  \\ \hline
\end{tabular}
\caption{ Table of the vanishing rates of outage probabilities  conditioned
on the  number of relay nodes available to forward, denoted by
$|\mathcal{D}(s)|$.  The  
multiplexing gain is denoted by $0 \leq r \leq 1/2$. The acronyms are defined
as: S-STC, Synchronous Space-Time Coded scheme
(Section~\ref{section:
syn. distc});  ICB-DD, Independent Coding Based Distributed Delay diversity
(Section~\ref{sec: indep. delay diversity}); RCB-DD, Repetition Coding Based
 Distributed Delay diversity (Section~\ref{sec: repetition delay diversity});
ICB-DD-L, Independent
Coding Based Distributed Delay diversity with Linearly modulated
waveforms(Section~\ref{linearly modulated Delay-D}); A-STC, Asynchronous
Space-Time Coded scheme (Section~\ref{section: asyn. coding}).
}
\label{results table}
\end{table*}

\begin{figure}
\begin{center}
\includegraphics[scale=0.45]{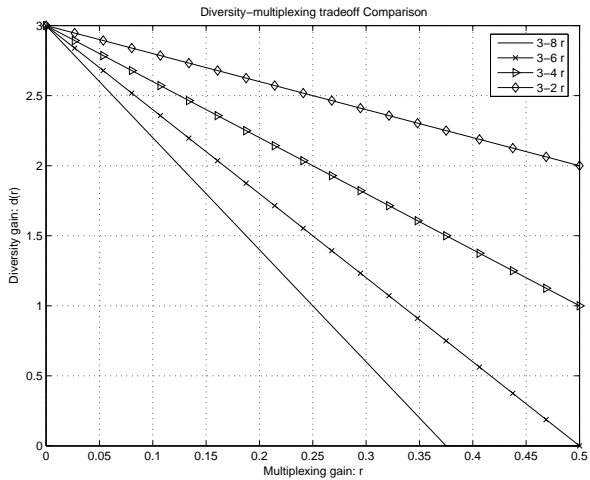}
\caption{Numerical Comparison of DM-tradeoff functions listed in
Table~\ref{results table} for $0 \leq r \leq 1/2$ and $\Delta_1 = 3/4$.
}\label{DM_comparison}
\end{center}
\end{figure}

\section{Conclusions}
\label{section: conclusion}

In this paper, we first show the lower-bound of the diversity-multiplexing
tradeoff developed by \cite{laneman_03} is actually the exact value
for a synchronous space-time coded cooperative diversity scheme. We
then propose two asynchronous cooperative diversity schemes, namely,
independent coding based distributed delay diversity and asynchronous
space-time coded relaying schemes. In terms of the overall DM-tradeoff,
both of them achieve the same performance as the synchronous one,
which demonstrates even at the presence of unavoidable asynchronism
between relay nodes, we don't loose diversity. Moreover, when all relay
nodes succeed in decoding the source information, the asynchronous
space-time coded approach achieves a better DM-tradeoff than the
synchronous scheme does and performs equivalently to transmitting
information through a parallel fading channel as far as the diversity
order is concerned.  Table~\ref{results table} summarizes the results
regarding the slope of conditional outage probability with respect to high
$\mbox{SNR}$ given $0 \leq |\mathcal{D}(s)| \leq 2$ number of relay nodes
available to forward.  The acronyms are defined as: S-STC, Synchronous
Space-Time Coded scheme (Section~\ref{section: syn. distc});  ICB-DD,
Independent Coding Based Distributed Delay diversity (Section~\ref{sec:
indep. delay diversity}); RCB-DD, Repetition Coding Based Distributed
Delay diversity (Section~\ref{sec: repetition delay diversity});
ICB-DD-L, Independent Coding Based Distributed Delay diversity with
Linearly modulated waveforms(Section~\ref{linearly modulated Delay-D});
A-STC, Asynchronous Space-Time Coded scheme (Section~\ref{section:
asyn. coding}).  Figure~\ref{DM_comparison} provides a comparison of
slope functions listed in Table~\ref{results table}.

In analyzing the asymptotic performance  of various approaches, a
bottleneck on the overall DM-tradeoff in relay channels is identified.
It is caused by restricting sources  transmitting only in the first
phase and relay nodes to employing   decode-and-forward strategy. A
simple mixing strategy is proposed to address this issue. By comparing
it with the NAF and DDF proposed by \cite{gamal_submitted}, we show the
mixing strategy achieves higher diversity gain than both DDF and NAF
over certain range of the multiplexing gain $r$ even though we still
let  source transmit only half of an entire frame.

As observed in Section~\ref{section: asyn. coding}, employing
properly designed $s(t)$ of a finite duration $T_s$ can even lead
to higher mutual information than synchronous space-time codes for
any $\mbox{SNR}$. This reveals the advantage of fully exploiting
both spatial and temporal  degrees of freedom in MIMO systems by employing
asynchronous space-time codes even in a  frequency non-selective
fading channel. The design of $s(t)$ and asynchronous space-time
codes, as well as the corresponding performance analysis is beyond
the scope of this paper and will be addressed  in our future work.

\appendix

\section{Appendix}

\subsection{Proof of Lemma~\ref{lemma1}}
\label{pf: lemma1}

\begin{proof}

For each subset $A_i$ of $\tilde{\mathcal A} =
\bigcup_{i=1}^{4} \tilde{\mathcal A}_i$ as defined in
Section~\ref{section: syn. distc}, we calculate the corresponding integrals
in (\ref{eq: outage prob.
without exp with one relay node})
 individually.

\no Over $\tilde{\mathcal
A}_1 = \left\{ \beta_{S,D} \geq 1, \beta_{R_1,D} \geq 1 \right\}$, we have
\begin{gather} \label{eq: first  of four terms for
the condit. one relay node}
\int_{\beta_{i,D} \in \tilde{\mathcal
A_1}}  \left( \log \widetilde{\mbox{SNR}} \right)^2 \prod_{k \in
\{S, R_1 \}} \widetilde{\mbox{SNR}}^{- \beta_{k,D}}
\tilde{\lambda}_{k,D} \, d \beta_{k,D} = \nonumber \\ 
 \prod_{k \in \{S, R_1 \}}
\tilde{\lambda}_{k,D}
\left[ \int_1^\infty \widetilde{\mbox{SNR}}^{-\alpha}
\left( \log\widetilde{\mbox{SNR}}\right)
\, d \alpha \right]^2  \nonumber \\
 =  \prod_{k \in \{S, R_1 \}}
\tilde{\lambda}_{k,D} \frac{1}{\widetilde{\mbox{SNR}}^2}.
\end{gather}

\no Over $\tilde{\mathcal A}_2 = \left\{ \beta_{S,D} \geq 1, \right.$ $\left. 1- 2r <
\beta_{R_1,D} < 1 \right\}$ or
$\tilde{\mathcal A}_3 = \left\{ 1-2r
< \beta_{S,D}  < 1, \beta_{R_1,D} \geq  1 \right\}$
the integral is
\begin{gather} \label{eq: second  and three of four terms for the
condit. one relay node}
\int_{\beta_{i,D} \in \tilde{\mathcal
A_i}}  \left( \log \widetilde{\mbox{SNR}} \right)^2 \prod_{k \in
\{S, R_1 \}} \widetilde{\mbox{SNR}}^{- \beta_{k,D}}
\tilde{\lambda}_{k,D} \, d \beta_{k,D}
 = \nonumber \\   \prod_{k \in \{S, R_1 \}}
\tilde{\lambda}_{k,D}
\int_1^\infty \int_{1-2r}^{1} \widetilde{\mbox{SNR}}^{-(\alpha_1+\alpha_2)}
 \cdot
\left( \log\widetilde{\mbox{SNR}}\right)^2
\, d \alpha_2 \, d \alpha_1  \nonumber \\
  =   \prod_{k \in \{S, R_1 \}}
\tilde{\lambda}_{k,D}
\frac{\left(\widetilde{\mbox{SNR}}\right)^{2r}-1}
{\widetilde{\mbox{SNR}}^2}, \, i =2, 3.
\end{gather}

\no Over $\tilde{\mathcal A}_4 = \left\{ 0 \leq \beta_{k, D} < 1,
\right.$ $\left. \sum_{k \in \{S, R_1 \}}\right.$ $\left. \beta_k > 2 -2r
\right\}$, we obtain
\begin{gather} \label{eq: four of four terms for the condit. one relay
node} \int_{\beta_{i,D} \in \tilde{\mathcal A_4}}  \left( \log
\widetilde{\mbox{SNR}} \right)^2 \prod_{k \in \{S, R_1 \}}
\widetilde{\mbox{SNR}}^{- \beta_{k,D}} \tilde{\lambda}_{k,D} \, d
\beta_{k,D}
 = \nonumber \\
 \prod_{k \in \{S, R_1 \}}
\tilde{\lambda}_{k,D}
\int_{1-2r} ^1 \int_{2-2r-\alpha_1}^{1} \widetilde{\mbox{SNR}}^{-(\alpha_1+\alpha_2)}
\cdot
\left( \log\widetilde{\mbox{SNR}}\right)^2
\, d \alpha_2 \, d \alpha_1  \nonumber \\
 =  \prod_{k \in \{S, R_1 \}} \tilde{\lambda}_{k,D}
\left[
\left( 2r \log \widetilde{\mbox{SNR}} -1 \right) \cdot \right. \nonumber \\
 \left. \left( \widetilde{\mbox{SNR}} \right)^{-(2-2r)} - \widetilde{\mbox{SNR}}^{-2} \right].
\end{gather}

Combining (\ref{eq: first  of four terms for the condit. one
relay node})-(\ref{eq: four of four terms for the condit. one
relay node}), we obtain the RHS of
(\ref{eq: asymp. conditional final of single realy node})
\begin{gather}
\int_{\beta_{i,D} \in \tilde{\mathcal A}}  \left( \log \widetilde{\mbox{SNR}} \right)^2
\prod_{k \in \{S, R_1 \}} \widetilde{\mbox{SNR}}^{- \beta_{k,D}}
\tilde{\lambda}_{k,D} \, d \beta_{k,D}
 =  \nonumber \\
\prod_{k \in \{S, R_1 \}} \tilde{\lambda}_{k,D} \left( 1 +  2 r \log
\widetilde{\mbox{SNR}} \right)  \left( \widetilde{\mbox{SNR}}
\right)^{-(2-2r)}
\nonumber \\
  \sim    \left( 2 r \log \widetilde{\mbox{SNR}}\right) \left(
\widetilde{\mbox{SNR}} \right)^{-(2-2r)} \prod_{k \in \{S, R_1 \}}
\tilde{\lambda}_{k,D},
\end{gather}

\no which completes the proof of  Lemma~\ref{lemma1}.

\end{proof}
\subsection{Proof of Lemma~\ref{lemma3}}
\label{lemma3proof}

\begin{proof}

To derive the asymptotic equivalence of
$\mbox{Pr} \left[ I_{stc} < R, \; |{\mathcal D}(s)| =2   \right]$,
  WLOG,
assume $\tilde{\lambda}_{R_1,D} > \tilde{\lambda}_{R_2,D}$ and denote $y = \sum_{k \in {\mathcal
D}(s) } |\tilde{\alpha}_{k,D}|^2 $.
The
probability density function (pdf) of $y$ is
\[
p(y) = \frac{\tilde{\lambda}_{R_1,D} \tilde{\lambda}_{R_2,D} }
{\tilde{\lambda}_{R_1,D}-\tilde{\lambda}_{R_2,D}} \left(
e^{-\tilde{\lambda}_{R_2,D} y}- e^{-\tilde{\lambda}_{R_1,D} y}
\right), \; y \geq 0.
\]

\no Define a normalized random variable $\beta_{R,D} = -\frac{\log  y}{\log
{
\widetilde{\mbox{SNR}}}}$ whose pdf is
\begin{gather} \label{ eq: pdf of the sum of two exp. rv}
p(\beta_{R,D})  =
\frac{\tilde{\lambda}_{R_1,D} \tilde{\lambda}_{R_2,D} }
{\tilde{\lambda}_{R_1,D}-\tilde{\lambda}_{R_2,D}}
\exp\left\{ -\lambda_{R_2,D} \widetilde{\mbox{SNR}}^{-\beta_{R,D}}\right\} \cdot \nonumber \\
   \left[ 1- \exp \left\{
-\left(\tilde{\lambda}_{R_1,D}-\tilde{\lambda}_{R_2,D} \right)
\widetilde{\mbox{SNR}}^{-\beta_{R,D}}  \right\}\right] \cdot
 \left(\log \widetilde{\mbox{SNR}} \right) \widetilde{\mbox{SNR}}^{-\beta_{R,D}} \nonumber \\
 \sim  \tilde{\lambda}_{R_1,D} \tilde{\lambda}_{R_2,D}
\left(\log \widetilde{\mbox{SNR}} \right)
\widetilde{\mbox{SNR}}^{-2\beta_{R,D}}, \end{gather}

\no for large $ \widetilde{\mbox{SNR}}$ and $\beta_{R,D} \geq 0$.
The conditional outage probability given two relay nodes are both in
the decoding set ${\mathcal D}(s)$ is
\begin{gather} \label{eq: condi. outage
given two relay nodes} \mbox{Pr} \left[ I_{stc} < R | |{\mathcal
D}(s)| =2   \right] \sim  
\int_{\beta_{i,D} \in \hat{\mathcal A}}
\left( \log \widetilde{\mbox{SNR}} \right)^2 \nonumber \\
\widetilde{\mbox{SNR}}^{- \beta_{S, D} - 2 \beta_{R,D}} \prod_{k
\in \left\{ S,R_1,R_2 \right\}} \tilde{\lambda}_{k,D} \, d
\beta_{S,D} d \beta_{R, D}, \end{gather}

\no where
\[
\hat{\mathcal A} = \left\{\underline{\beta}:  {\sum_{i \in
\left\{S,R \right\}} (1- \beta_{i,D})^{+}} < 2r , \beta_{i,D} \geq
0 \right\}.\]

\no By employing the same method as the one through which
(\ref{eq: asymp. conditional final of single realy node}) is
obtained, it can be  shown that 
\begin{gather} \label{eq: whole link conditional outage with 2 relay nodes} 
\mbox{Pr} \left[ I_{stc} <
R | |{\mathcal D}(s)| =2   \right]  \sim  \nonumber \\ 
2 \prod_{k \in \left\{
S,R_1,R_2 \right\}} \tilde{\lambda}_{k,D} \left(
\widetilde{\mbox{SNR}} \right)^{-3 + 4 r} \left[ 1- \frac{1}{2}
\left( \widetilde{\mbox{SNR}} \right)^{-2r} \right] \nonumber \\
 \sim  2 \prod_{k \in \left\{ S,R_1,R_2 \right\}}
\tilde{\lambda}_{k,D} \left( \widetilde{\mbox{SNR}} \right)^{-3 +
4 r} . \end{gather}

\no As for the probability of $|\mathcal{D}(s)|=2$,  we have
$\mbox{Pr} \left[ \left| {\mathcal D}(s) \right| = 2
\right] \sim 1 $ resulting from  (\ref{eq: asymptotic of decoding set}). Thus,
the overall conditional outage probability is
\[
\mbox{Pr} \left[ I_{stc} < R, \; |{\mathcal D}(s)| =2   \right]
\sim
 2 \prod_{k \in \left\{ S,R_1,R_2 \right\}} \tilde{\lambda}_{k,D}
\left( \widetilde{\mbox{SNR}} \right)^{-3 + 4 r},
\]

\no which completes the proof of  Lemma~\ref{lemma3}.

\end{proof}
\subsection{Proof of Theorem~\ref{linearly modulated TDA}}
\label{theoremLTDAProof}

\begin{proof}

Given $|\mathcal{D}(s)|=2$,
the canonical receiver for the resulting equivalent  $2$-path  fading
channel consists of a whitened matched filter (WMF)  and  a symbol
rate sampler \cite{cioffi_1995}.
The Fourier transform of the  impulse
response  of this  equivalent channel is
 $F(f) = H(f) S(f)$, where $H(f) =
\sum_{k} \alpha_{R_k,D} e^{-j2 \pi f \tau_k}$ and $S(f)$ is the
Fourier transform of $s(t)$.  The mutual information of this
$2$-path fading  channel given $\{\alpha_{R_k,D} = r_k e^{j\theta_k} \}$ is
\cite[pp. 2597]{cioffi_1995}
\be \label{eq: capacity of multipath channel with WMF}
I_{2-TDA} = \frac{1}{2 \pi} \int_{-\pi}^{\pi} \log \left[
1 +  \rho_0 |S_{hh} (\omega )|^2 \right] \, d \omega, \ee

\no where
$|S_{hh} (\omega )|^2 = \sum_{k} h(k) e^{j k \omega}$
is  the discrete
Fourier transform of
$h(k)$, which is  the sampling output of the matched
filter for $F(t) = s(t) \alpha_{R_1,D} + s(t-\tau) \alpha_{R_2,D}$, i.e.
\[
h(k)  = \int_{-\infty}^{\infty} F(t) F^*(t -k T_s ) \, dt,  \]

\no with  $\tau = \tau_2 -\tau_1 $ denoted as the relative delay. WLOG, we
assume $\tau \in (0, T_s]$ \cite{roger_91}.
 Due to the
time-limited  constraint on $s(t)$, we obtain
 $h_{k}=0$ for $|k| \geq 2$, and
\beeq h(0) = |\alpha_{R_1,D}|^2 +  |\alpha_{R_2,D}|^2 + \rho_{12}
\left(
 \alpha_{R_1,D} \alpha_{R_2,D}^{*} +  \alpha_{R_2,D}
\alpha_{R_1,D}^{*} \right) \eeeq

\no and \beeq h(1) =  \alpha_{R_2,D} \alpha_{R_1,D}^{*}  \rho_{21},
\,  h(-1) = \alpha_{R_1,D} \alpha_{R_2,D}^{*}  \rho_{21} , \eeeq

\no where $\rho_{12}$ and $\rho_{21}$ are correlation coefficients
of $s(t)$ determined by $\rho_{12} = \int_{0}^{T_s} s(t) s(t-
\tau) \, dt$ and $\rho_{21} =
\int_{0}^{T_s} s(t) s(t +T_s- \tau) \, dt$.

From Cauchy Schwartz inequality and $\int_{0}^{T_s}
|s(t)|^2 \, dt =1$,  we have $|\rho_{12}|
<  1$ and  $|\rho_{21}| \leq 1$ for $\tau \in (0, T_s]$,  and
\begin{gather} \label{eq: upperbound of rho}
|\rho_{12}| + |\rho_{21}|  = \nonumber \\ \left| \int_0^{T_s}
s(t) s(t-\tau)\, dt \right| + \left| \int_0^{T_s}
s(t) s(t + T_s - \tau)\, dt \right| \nonumber \\
 \leq   \int_{0}^{T_s} |s(t)| \left( | s(t -\tau)|+ | s(t+T_s -\tau)|\right)
\, dt \leq 
\nonumber \\
   \left[ \int_0^{T_s} |s(t)|^2 \,dt \right]^{1/2}
\left[ \int_0^{T_s} \left( \left| s(t-\tau) \right| + \left| s(t+T_s -\tau)
\right| \right)^2
\, dt
\right]^{1/2} \nonumber \\
 =  1.
  \end{gather}

\no Substituting $h(k)$ into $|S_{hh} (\omega )|^2 $ yields
\begin{gather} \label{eq: I-{2-TDA}} 
I_{2-TDA} =  \nonumber \\   \frac{1}{2 \pi}
\int_{-\pi}^{\pi} \log \left[ 1 +  \rho_0 \left(  |\alpha_{R_1,D}|^2
+  |\alpha_{R_2,D}|^2 +  \alpha_{R_1,D} \alpha_{R_2,D}^{*} \right. \right.
\nonumber \\
\left. \left. ( \rho_{12} + \rho_{21} e^{j\omega}) 
   \alpha^*_{R_1,D} \alpha_{R_2,D} ( \rho_{12} +
\rho_{21} e^{-j\omega})
\right) \right] \, d \omega  \nonumber \\
 =   \frac{1}{2 \pi} \int_{-\pi}^{\pi} \log \left[ 1+ \rho_0
\left(\left| \alpha_{R_1,D} + \alpha_{R_2,D}(\rho_{12} +
\rho_{21}e^{-j\omega} ) \right|^2  \right. \right. \nonumber \\
  \left. \left. + |\alpha_{R_2,D}|^2 \left(1- \left| \rho_{12} +
\rho_{21}e^{-j\omega} \right|^2 \right)
  \right)  \right] \, d \omega\nonumber \\
 =  \frac{1}{2 \pi} \int_{-\pi}^{\pi}  \log \left[
1+ a + b \cos (\theta_1 - \theta_2 + \omega) \right]\, d \omega \nonumber \\
 =  \log \left[ 1+a + \sqrt{(1+a)^2 -b^2} \right] -1
\end{gather}

\no where the last equality is from Eq. (\ref{eq: table
integral}) with  $a$ and $b$ defined as:
\be a &  = & \rho_0
\left(|r_1|^2 + |r_2|^2 + 2 \rho_{12} r_1 r_2 \cos(\theta_1
-\theta_2 ) \right)
\nonumber \\
& = &  \rho_0 \left( \left|\alpha_{R_1,D} + \alpha_{R_2,D} \rho_{12}
\right|^2 + |\alpha_{R_2,D}|^2 ( 1- |\rho_{12}|^2) \right) \nonumber \\
b & = & 2  \rho_{21} r_1 r_2 \rho_0.
\ee

\no   Given
 $|\rho_{12}| + |\rho_{21}|\leq 1$, it can be shown
$ a \geq  b$ and $a \geq 0$ which enables us to bound $I_{2-TDA}$ in (\ref{eq:
I-{2-TDA}})
by \beeq \label{eq:  bound of I-{2-TDA}} I_{2-TDA}^{(L)}
\stackrel{\bigtriangleup}{=} \log \left[ 1+a \right] -1 \leq
I_{2-TDA} \leq I_{2-TDA}^{(U)} \stackrel{\bigtriangleup}{=}
\log \left[ 1+a \right]
 . \eeeq

\no Define random variables $X_1= \alpha_{R_1,D} + \alpha_{R_2,D}
\rho_{12} $ and $X_2 = \alpha_{R_2,D} \sqrt{1- |\rho_{12}|^2}$.  We can then 
rewrite $ I_{2-TDA}^{(U)} = \log\left[ 1+ \rho_0
(|X_1|^2 + |X_2|^2 ) \right]$ and $ I_{2-TDA}^{(L)} = \log\left[
1+ \rho_0 (|X_1|^2 + |X_2|^2 ) \right]-1$. Clearly, the vector
$[X_1, X_2]'$ is a linear transformation of the random  vector
$[\alpha_{R_1,D}, \alpha_{R_2,D}]'$, i.e. 
\be \left[
\begin{array}{l} X_1 \\ X_2 \end{array} \right] & = & \left[
\begin{array}{ll}
1 & \rho_{12} \\
0 &  \sqrt{1- |\rho_{12}|^2}
\end{array}
\right]
\left[ \begin{array}{l}
\alpha_{R_1,D} \\
\alpha_{R_2,D}
\end{array}
\right] = {\bf B}  \left[ \begin{array}{l}
\alpha_{R_1,D} \\
\alpha_{R_2,D}
\end{array}
\right] \nonumber \\ & = &  {\bf B}  \left[ \begin{array}{ll}
\sigma_{R_1,D} & 0 \\
0 & \sigma_{R_2,D} \end{array}
\right]
 \left[ \begin{array}{l}
\hat{\alpha}_{R_1,D} \\
\hat{\alpha}_{R_2,D}
\end{array}
\right],  
\ee

\no  where 
\[
B = \left[
\begin{array}{ll}
1 & \rho_{12} \\
0 &  \sqrt{1- |\rho_{12}|^2}
\end{array}
\right],
\]

\no and the entries 
of $\left[\hat{\alpha}_{R_1,D},
\hat{\alpha}_{R_2,D} \right]^{'}$  are i.i.d.   complex
Gaussian random variables with zero mean and unit variance.
Define a upper-triangle matrix
\[ {\bf A} =  {\bf B}  \left[ \begin{array}{ll}
\sigma_{R_1,D} & 0 \\
0 & \sigma_{R_2,D} \end{array}
\right].
\]

\no The matrix ${\bf A}$ can therefore  be
decomposed  as ${\bf A} = {\bf U} {\bf D_A} {\bf U}^{\dagger}$ using singular
value decomposition,  where ${\bf U}$ is a unitary matrix and ${\bf D_A} =
\mbox{diag}\left[\sigma_{R_1,D}, \sigma_{R_2,D}\cdot \sqrt{1-|\rho_{12}|^2}
\right]$
 is a diagonal matrix
whose diagonal entries are the eigenvalues of the upper-triangular
matrix ${\bf A}$. Decomposing ${\bf A}$ as such, we obtain
$|X_1|^2 + |X_2|^2 = |\tilde{\alpha}_{R_1}|^2 +
|\tilde{\alpha}_{R_2}|^2 \sqrt{1- |\rho_{12}|^2} $, where
$[\tilde{\alpha}_{R_1}, \tilde{\alpha}_{R_2}]'$ is a vector having the
same joint distribution as  $[ \alpha_{R_1, D}, \alpha_{R_2,D}]$.
Given the  bounds on $I_{2-TDA}$, the overall mutual information
$I_{L-TDA}$ can be bounded accordingly as $I_{L-TDA} \in [
I_{L-TDA}^{(L)}, I_{L-TDA}^{(U)}]$, where $I_{L-TDA}^{(L)}$ and
$I_{L-TDA}^{(U)}$ are 
\begin{gather} I_{L-TDA}^{(L)} = \frac{1}{2} \log
\left( 1 + \rho_0 |\alpha_{S,D}|^2 \right) + \nonumber \\
\frac{1}{2} \log
\left[ 1 + \rho_0 |\tilde{\alpha}_{R_1}|^2 + \rho_0
|\tilde{\alpha}_{R_2}|^2 \sqrt{1- |\rho_{12}|^2} \right]-1
 \end{gather}

\no and \begin{gather} I_{L-TDA}^{(U)} = \frac{1}{2} \log \left( 1 + \rho_0
|\alpha_{S,D}|^2 \right)+ \nonumber \\
 \frac{1}{2} \log \left[ 1 + \rho_0
|\tilde{\alpha}_{R_1}|^2 + \rho_0 |\tilde{\alpha}_{R_2}|^2
\sqrt{1- |\rho_{12}|^2} \right].
\end{gather}

As shown previously,  given $\int_{0}^{T_s} |s(t)|^2 \, dt =1$, for any
$\tau \in (0,T_s]$, we have $|\rho_{12}|<1$ and thus $1-
|\rho_{12}|^2 >0 $ which implies the asymptotic behavior of outage
probabilities $\mbox{Pr} \left[ I_{L-TDA}^{(U)}<R,
|\mathcal{D}(s)|=2 \right]$ and $\mbox{Pr} \left[
I_{L-TDA}^{(L)}<R, |\mathcal{D}(s)|=2 \right]$ is similar as the
one characterized by Lemma~\ref{lemma3} for synchronous space-time
coded cooperative diversity scheme. Therefore, applying the same techniques in
proving Lemma~\ref{lemma3} yields
(\ref{eq: linearly modulated TDA}) and thus Theorem~\ref{linearly
modulated TDA} is proved.

\end{proof}
\subsection{Proof of Theorem~\ref{theorem: positive defininity}}
\label{positiveproof}

\begin{proof}

Denote $\underline{S}(t) =[s(t), s(t-\tau)]^T$ and
$\underline{S}_w(t)= \sum_{k=-2}^2
\underline{S}(t-kT_s)e^{jk\omega}$, where $s(t) =0, t \notin [0,
2T_s]$. The matrix ${\bf \tilde{T}_E}(\omega)$ defined in
(\ref{eq: Original h matrix}) is : 
\begin{gather} \label{eq: expansion of H(omega) matrix} 
{\bf \tilde{T}_E}(\omega)  =  \sum_{k=-2}^2
{\bf H_E} (k) e^{-j k \omega}
 =   \int_{-\infty}^{\infty} \underline{S}(t) \underline{S}^{\dagger}_w (t)\, dt \nonumber \\
 =  \int_{0}^{3 T_s} \underline{S}(t)
\underline{S}^{\dagger}(t)\, dt + e^{-j2\omega}
\int_{0}^{T_s}  \underline{S}(t) \underline{S}^{\dagger}(t+2 T_s)\, dt \nonumber \\
   + e^{j2\omega} \int_{2 T_s}^{3 T_s}  \underline{S}(t)
\underline{S}^{\dagger}(t-2T_s)\, dt  + e^{j \omega} \int_{T_s}^{3
T_s}  \underline{S}(t) \underline{S}^{\dagger}(t-T_s)
 \, dt  \nonumber \\
 + e^{-j \omega} \int_{0}^{2 T_s}  \underline{S}(t)
\underline{S}^{\dagger}(t+T_s)  \, dt
\nonumber \\
 =  \int_{0}^{T_s} \left[ \sum_{k=0}^2 \underline{S}(t+kT_s)
e^{j k\omega} \right] \left[ \sum_{k=0}^2 \underline{S}(t+kT_s)
e^{j k\omega} \right]^{\dagger} \, dt, \end{gather}

\no where the above equations are derived by exploiting the finite
duration of $s(t)$, as well as the definition of parameters in
(\ref{eq: coefficients a_1 and d_1})-(\ref{eq: coeef. f_1 c_2}).
As implied by the last equation in (\ref{eq: expansion of H(omega)
matrix}), ${\bf \tilde{T}_E}(\omega) $ is a non-negative definite
matrix. This result can be extended in a similar manner to the
case when $s(t)$ spans over any arbitrary  finite $M T_s$ periods
where $M \geq 1$ is an integer, i.e. $s(t) =0, t \notin [0, M
T_s]$. Define $\underline{S}^{(M)}_w(t)= \sum_{k=-M}^M
\underline{S}(t-kT_s)e^{jk\omega}$. We obtain
 \begin{gather} \label{eq: extension of H(omega)} 
{\bf \tilde{T}_E}^{(M)}(\omega)  = 
\int_{-\infty}^{\infty} \underline{S}(t) \left(
\underline{S}^{(M)}_w (t)
\right)^{\dagger}\, dt \nonumber \\
 =  \int_{0}^{T_s} \left[ \sum_{k=0}^M \underline{S}(t+kT_s)
e^{j k\omega} \right] \left[ \sum_{k=0}^M \underline{S}(t+kT_s)
e^{j k\omega} \right]^{\dagger} \, dt ,\end{gather}

\no which is a non-negative definite matrix for $M \geq 1$. Define
$F_1(t,\omega) =  \sum_{k=0}^M s(t+kT_s) e^{j k\omega}$ and
$F_2(t,\omega) =  \sum_{k=0}^M s(t-\tau+kT_s) e^{j k\omega}$ for
all $t \in [0, T_s]$ and $\omega \in [-\pi, \pi]$. For a given
$t$, $F_1(t,\omega)$ and $F_2(t,\omega)$ are the discrete time
Fourier transforms of sampled signals of $s(t)$ and $s(t-\tau)$ at
time instants $\left\{ t+kT_s, k=0, 1, \cdots, M \right\}$,
respectively.  If there exists a non-zero complex vector $
\underline{b} = [b_0, b_1]$ such that $ \underline{b} {\bf
\tilde{T}_E}^{(M)}(\omega) \underline{b}^{\dagger} =0$ for some
$\omega$, it indicates ${\bf \tilde{T}_E}^{(M)}(\omega)$ has a
zero eigenvalue for the specified $\omega$, and thus  we must have
the following linear relationship associated with $F_j(t,
\omega)$: $ b_0 F_1(t, \omega) + b_1 F_2(t, \omega)=0$ for any $t
\in [0, T_s]$. Therefore,  if $s(t)$ is chosen to make $F_1(t,
\omega)$ and $F_2(t, \omega)$ linearly independent with respect to
$t$ for any given $\omega$, ${\bf \tilde{T}_E}^{(M)}(\omega)$ is
always positive definite satisfying  $\underline{b} {\bf
\tilde{T}_E}^{(M)}(\omega) \underline{b}^{\dagger}  >0$ for any
non-zero  $\underline{b}$ and $\forall \omega \in [-\pi, \pi]$.
Let 
\begin{gather}  G(\underline{b}, \omega)  =  \underline{b} {\bf
\tilde{T}_E}^{(M)}(\omega)
\underline{b}^{\dagger} \nonumber \\
 =  \int_0^{T_s} \left| b_0 F_1(t,\omega) + b_1 F_2 (t, \omega)
\right|^2 \, dt, \, ||\underline{b}||=1, \end{gather}

\no denote a continuous function of $\underline{b}$ and $\omega$
defined over a closed and bounded region, where
$||\underline{b}||$ is the Euclidean norm of $\underline{ b}$.
Define
\[\lambda^{(M)}_{\mbox{min}} = \inf_{\omega\in[0, 2\pi], ||\underline{b}||=1}
G(\underline{b}, \omega), \, \lambda^{(M)}_{\mbox{max}} =
\sup_{\omega\in[0, 2\pi], ||\underline{b}||=1}G(\underline{b},
\omega)
\]
\no  By Weierstrass' Theorem \cite[pp. 654]{dimitri_nonlinear},
the greatest lower bound $\lambda^{(M)}_{\mbox{min}}$ and least
upper bound  $\lambda^{(M)}_{\mbox{max}}$ of $G(\underline{b},
\omega)$ is attainable. Therefore, if  $s(t)$ is properly selected
as specified above which results  in positive definite matrices
${\bf \tilde{T}_E}^{(M)}(\omega)$, $\lambda^{(M)}_{\mbox{min}}$
and $\lambda^{(M)}_{\mbox{max}}$ are achievable and both of them
are positive. In addition, $\lambda^{(M)}_{\mbox{max}} $ can be
further upper-bounded by some finite constant  as shown below: 
\begin{gather}
\label{eq: upperbound of max eigenvalue}
\lambda^{(M)}_{\mbox{max}}(\omega)  = 
  \sup_{ ||\underline{b}||=1} \underline{b} {\bf \tilde{T}_E}^{(M)}(\omega)  \underline{b}^{\dagger}
\nonumber \\
  \leq   \mbox{Tr} \left( {\bf
\tilde{T}_E}^{(M)}(\omega) \right) =   2 \sum_{k=-M}^M \int_{-\infty}^{\infty}
s(t) s(t-kT_s) e^{-j k \omega} \, dt \nonumber \\
 \leq   2 \sum_{k=-M}^M  \left|\int_{-\infty}^{\infty}
s(t) s(t-kT_s) \, dt \right|   \nonumber \\
 \leq   2 \sum_{k=-M}^M \left[\int_{-\infty}^{\infty} |s(t)|^2 \, dt \right]^{1/2}
 \left[\int_{-\infty}^{\infty} |s(t-kT_s)|^2 \, dt \right]^{1/2} \nonumber \\
 =  2 (2 M+1), \, \omega \in [0, 2 \pi],  \end{gather}

\no where the first and second  inequalities are  due to  the
positive definiteness of $ {\bf \tilde{T}_E}^{(M)}(\omega) $, and
Cauchy-Schwartz inequality yields the third inequality. The last
equality is because $s(t)$ has unit energy.

When $M=2$, this proves  Theorem~\ref{theorem: positive
defininity}.
 When $M=1$, i.e. the waveform $s(t)$ is confined within one
symbol interval, the condition stated in \cite[pp. 4]{roger_92} is
a special case of our result which reduces  to the following
condition for $M=1$: $s(t)$ and $s(t-\tau)+
s(t-\tau+T_s)e^{j\omega}$ are linearly independent with respect to
$t\in [0, T_s]$ which is equivalent to
 $ s(t)$ and $s(t+T_s-\tau)e^{j \omega}$, as well as $s(t)$ and $s(t-\tau)$ are linearly independent
over $t \in [0, \tau]$ and $t \in (\tau, T_s]$, respectively.
Also, we can observe from the second equality in (\ref{eq:
upperbound of max eigenvalue}) that $\mbox{Tr} \left( {\bf
\tilde{T}_E}^{(M)}(\omega) \right) =  2 \int_{0}^{T_s} |s(t)|^2  \,
dt =2$ in this case. This fact will be exploited when we compare
the mutual information of a MISO channel using asynchronous
space-time codes with that employing synchronous space-time codes.

Actually, the condition under which ${\bf
\tilde{T}_E}^{(1)}(\omega) $ is positive definite can be further
exposed by looking more closely at the parameters defined in
(\ref{eq: coefficients a_1 and d_1})-(\ref{eq: coeef. f_1 c_2})
for $s(t) =0, \, t \notin [0, T_s]$. In this case, it is
straightforward to show that $a_1= d_1 = c_2 = f_1 =0$ for $\tau_2
\geq \tau_1$. Therefore, the product of eigenvalues of the
Hermitian matrix ${\bf \tilde{T}_E}^{(1)}(\omega)$ is
\begin{gather} \left(
1 + 2 a_1 \cos \omega \right)^2 - \left| c_1 e^{-j \omega} + c_2
e^{-j 2\omega} + c_0 + f_1 e^{j \omega}\right|^2 \nonumber \\ =
 1- \left|c_0 + c_1 e^{-j\omega} \right|^2 \geq 0
\end{gather}

\no where the inequality can be shown as follows.
As defined in (\ref{eq: coefficients a_1 and d_1})-(\ref{eq: coeef. f_1 c_2}),
$c_0$ and $c_1$ are correlation coefficients
of $s(t)$ determined by $c_0= \int_{0}^{T_s} s(t) s(t-
\tau) \, dt $ and $c_1 =
\int_{0}^{T_s} s(t) s(t +T_s- \tau) \, dt$. From Cauchy Schwartz inequality and $\int_{0}^{T_s}
|s(t)|^2 \, dt =1$,
\begin{gather} \label{eq: upperbound of c_0}
|c_0 + c_1 e^{-j \omega}|  =  |c_0 + c_1 e^{j \omega}| \nonumber \\ 
= \left| \int_0^{T_s}
s(t) \left[ s(t-\tau) +
s(t+T_s-\tau)e^{j\omega} \right] \, dt \right| \leq \nonumber \\
 \left[ \int_0^{T_s} |s(t)|^2 \,dt \right]^{1/2}
\left[ \int_0^{T_s} \left| s(t-\tau) + s(t+T_s -\tau) e^{j\omega}\right|^2
\right]^{1/2} \nonumber \\
 =  1
\end{gather}

\no where the last equality is because $s(t-\tau)$
and $s(t+T_s-\tau)$ have no overlap over $ t \in [0, T_s]$, and
 the inequality becomes equality when  $s(t) = C
\left[s(t-\tau) + s(t + T_s - \tau) e^{j \omega} \right]$ where $|C|=1$ is a
 constant. This demonstrates only when
$s(t)$ and $s(t-\tau)+
s(t-\tau+T_s)e^{j\omega}$ are linearly independent with respect to
$t\in [0, T_s]$ for any $\omega \in [-\pi, \pi]$, can we have
a strict inequality in (\ref{eq: upperbound of c_0}) which agrees with
the condition on $s(t)$ in
Theorem~\ref{theorem: positive defininity} and thus
verifies it from another
perspective for $M=1$.

Note when the waveform $s(t)$ is a truncated version of a
squared-root-raised-cosine waveform \cite{proakis_95} spanning
over $M$ symbol intervals such that $\int_{0}^{( M+ |k|) T_s} s(t)
s(t-k T_s) \, dt \approx \delta_{k}$, where $\delta_{0} =1$ and
$\delta_{k}=0$ for $ k \neq 0$,  the sum of eigenvalues of the
matrix can be approximated as $\mbox{Tr} \left( {\bf
\tilde{T}_E}^{(M)}(\omega) \right) \approx 2$. Again, this
property will be exploited when we compare the mutual information
of two MIMO systems employing synchronous and asynchronous
space-time codes, respectively.

\end{proof}

\begin{centering}
{\bf Acknowledgment}
\end{centering}

The author would like to thank anonymous reviewers for their valuable
suggestions for improving the presentation of this paper.

\bibliographystyle{IEEEtran}

\begin{biography}
[
]
{Shuangqing Wei}

Shuangqing Wei graduated in Electrical Engineering from Tsinghua
University with BE and MS in 1995 and 1998 and then obtained his Ph.D
in EE from the University of Massachusetts, Amherst in 2003.  He is
currently an Assistant Professor in the Department of Electrical and
Computer Engineering  at Louisiana State University.  His areas of
interest are in communication theory, information theory, coding theory
and their applications to wireless networks.

\end{biography}

\end{document}